\pdfoutput=1 

\documentclass[11pt,a4paper]{article}

\usepackage{jheppub}
\usepackage{atlasphysics}

\usepackage{rotating}
\usepackage{epstopdf}
\usepackage{overpic}
\usepackage{subfigure}
\usepackage{array}
\usepackage{lineno}  
\usepackage[section]{placeins}
\usepackage{pdflscape}

\usepackage{preprintcover}  
\PreprintCoverPaperTitle{Search for the direct production of charginos, neutralinos and staus in final 
states with at least two hadronically decaying taus and missing transverse momentum in $pp$ collisions at $\sqrt{s}$~=~8\,TeV with the ATLAS detector}  
\PreprintIdNumber{CERN-PH-EP-2014-121}  
\PreprintCoverAbstract{
Results of a search for the electroweak associated production of charginos and next-to-lightest neutralinos, pairs of charginos or pairs of tau
 sleptons are presented. These processes are characterised by final states with at least  two hadronically decaying tau leptons,
missing transverse momentum and low jet activity. 
The analysis is based on an integrated luminosity of 
 20.3~fb$^{-1}$ of proton--proton collisions 
 at $\sqrt{s}=8$~TeV recorded with the ATLAS experiment at the Large Hadron Collider.
No significant excess is observed with respect to the predictions from Standard Model processes.
Limits are set at 95\% confidence level on the masses of the lighter chargino and next-to-lightest neutralino for various hypotheses
for the lightest neutralino mass in simplified models.
In the scenario of direct production of chargino pairs, with each chargino decaying into the lightest neutralino via an intermediate tau slepton, 
chargino masses up to 345 GeV are excluded for a 
massless lightest neutralino. For associated production of mass-degenerate charginos and next-to-lightest 
neutralinos, both decaying into the lightest neutralino via an intermediate tau slepton, masses up to 410 GeV are excluded for a massless lightest neutralino.
}  
\PreprintJournalName{JHEP}  

\graphicspath{{figures/}}

\newcommand{\alpgen}{{\tt ALPGEN}}
\newcommand{\powheg}{{\tt POWHEG}}

\newcommand{\mttwo}{\ensuremath{m_\mathrm{T2}}}

\makeatletter
\g@addto@macro\bfseries{\boldmath}
\makeatother

\newcommand{\gaugino}[2]{\ensuremath{\tilde\chi_{#1}^{#2}}}
\newcommand{\neutralino}[1]{\gaugino{#1}{0}}
\newcommand{\chargino}[1]{\gaugino{#1}{\pm}}
\newcommand{\Cone}{\chargino{1}}

\newcommand{\Ntwo}{\neutralino{2}}
\newcommand{\ConeCone}{\gaugino{1}{+}\gaugino{1}{-}}
\newcommand{\ConeNtwo}{\Cone\Ntwo}

\newcommand{\limsigvis}{\ensuremath{\sigma_\mathrm{vis}^{95}}} 

\newcolumntype{L}{>{$}l<{$}}
\newcolumntype{C}{>{$}c<{$}}
\newcolumntype{R}{>{$}r<{$}}

\begin{document}


\title{Search for the direct production of charginos, neutralinos and staus in final 
states with at least two hadronically decaying taus and missing transverse momentum in $pp$ collisions at $\sqrt{s}$~=~8\,TeV with the ATLAS detector}

\author{The ATLAS Collaboration}

\abstract{%
Results of a search for the electroweak associated production of charginos and next-to-lightest neutralinos, pairs of charginos or pairs of tau
 sleptons are presented. These processes are characterised by final states with at least  two hadronically decaying tau leptons,
missing transverse momentum and low jet activity. 
The analysis is based on an integrated luminosity of 20.3~fb$^{-1}$ of proton--proton collisions 
 at $\sqrt{s}=8$~TeV recorded with the ATLAS experiment at the Large Hadron Collider.
No significant excess is observed with respect to the predictions from Standard Model processes.
Limits are set at 95\% confidence level on the masses of the lighter chargino and next-to-lightest neutralino for various hypotheses
for the lightest neutralino mass in simplified models.
In the scenario of direct production of chargino pairs, with each chargino decaying into the lightest neutralino via an intermediate tau slepton, 
chargino masses up to 345 GeV are excluded for a 
massless lightest neutralino. For associated production of mass-degenerate charginos and next-to-lightest 
neutralinos, both decaying into the lightest neutralino via an intermediate tau slepton, masses up to 410 GeV are excluded for a massless lightest neutralino.
}

\maketitle


\section{Introduction}

Supersymmetry (SUSY)~\cite{Miyazawa:1966,Ramond:1971gb,Golfand:1971iw,Neveu:1971rx,Neveu:1971iv,Gervais:1971ji,Volkov:1973ix,Wess:1973kz,Wess:1974tw} 
is a promising extension of the Standard Model (SM) of particle physics. For each SM particle supersymmetry predicts 
the existence of a super-partner (also referred to as a `sparticle'), whose spin differs by one half
unit from the corresponding SM partner.
Supersymmetric theories provide elegant solutions to unanswered questions of the SM, such as the hierarchy 
problem~\cite{Weinberg:1975gm,Gildener:1976ai,Weinberg:1979bn,Susskind:1978ms}.
In $R$-parity-conserving SUSY models~\cite{Fayet:1976et,Fayet:1977yc,Farrar:1978xj,Fayet:1979sa,Dimopoulos:1981zb},
 SUSY particles are always produced in pairs and
the lightest supersymmetric particle (LSP) provides a dark matter candidate~\cite{DM,Goldberg:1983nd,Ellis:1983ew}. 

In SUSY models, the mass eigenstates formed from the linear superpositions of the SUSY partners of the
charged and neutral Higgs bosons and electroweak gauge bosons, the charginos ($\susy{\chi}^{\pm}_{i}$, $i$ = 1, 2)
and neutralinos ($\susy{\chi}^{0}_{j}$, $j$ = 1, 2, 3, 4 in the order of increasing masses), as well as the sleptons (superpartners
of the leptons,\footnote{
The sleptons are referred to as left- or right-handed ($\susy{\ell}_L$ or $\susy{\ell}_R$), depending on the helicity of their SM partners.
The slepton mass eigenstates are a mixture of $\susy{\ell}_L$ and $\susy{\ell}_R$ and labelled as 
$\susy{\ell}_1$ and $\susy{\ell}_2$.} $\susy{\ell}$ and $\susy{\nu}$) can be sufficiently light to be produced at the Large Hadron Collider (LHC)~\cite{LHC:2008}. 
Naturalness arguments suggest that the lightest third-generation sparticles, charginos and neutralinos should have masses of a few hundred 
GeV to protect the Higgs boson mass from quadratically divergent quantum corrections~\cite{Barbieri:1987fn,deCarlos:1993yy}.
Furthermore, light sleptons could play a role in the co-annihilation of neutralinos, leading to a dark matter relic density consistent with 
cosmological observations~\cite{Belanger:2004ag,King:2007vh}, and their mass is expected to be in the $\mathcal{O}$(100 GeV) range in gauge-mediated~\cite{Dine:1981gu,AlvarezGaume:1981wy,Nappi:1982hm,Dine:1993yw, Dine:1994vc,Dine:1995ag} 
and anomaly-mediated~\cite{Randall:1998uk,Giudice:1998xp} SUSY breaking scenarios. 
Models with light tau sleptons (the staus, labelled as $\tilde{\tau}$ in the following) are consistent 
 with current dark matter searches \cite{Vasquez:2011}.
  
This paper presents a search for electroweak production of charginos, next-to-lightest neutralinos and staus 
in events with at least two hadronically decaying tau leptons, missing transverse momentum and low jet activity, 
using the 2012 dataset of $\sqrt{s}=8$~TeV proton--proton collisions collected with the ATLAS detector.
Previous searches from the ATLAS collaboration cover electroweak production of supersymmetric particles in final states with 
electrons and muons~\cite{2Lep-2012} using signal models where the neutralinos and charginos decay with equal probability to all lepton flavours, 
and final states with exactly three leptons of any flavour (electrons, muons, and hadronically decaying taus)~\cite{3Lep-2012} using models similar to
those described in this paper. 
A search for associated production of charginos and next-to-lightest neutralinos in stau dominated scenarios has recently been published by the CMS collaboration \cite{Khachatryan:2014qwa}.
The combined LEP limits on the stau and chargino masses are
$m_{\stau}>87$-$93\GeV$ (depending on the ${\tilde{\chi}_{1}}^{0}$ mass) and
$m_{\chargino{1}}>103.5\GeV$~\cite{lepsusy,Abdallah:2003xe,Acciarri:1999km,Abbiendi:2003sc}.
It should be noted that the stau mass limit from LEP assumes gaugino mass unification, which is not assumed in the results presented here.

\section{SUSY scenarios}

SUSY scenarios characterised by the presence of light charginos, next-to-lightest neutralinos
and sleptons can be realised in the general framework of the phenomenological Minimal Supersymmetric Standard Model 
(pMSSM)~\cite{pmssm1,pmssm2,pmssm3}. The dominant processes are the electroweak production of $\susy{\chi}^{\pm}_{1}\susy{\chi}^{0}_{2}$ and $\susy{\chi}^{\pm}_{1}\susy{\chi}^{\mp}_{1}$,  such as $q\bar{q}\to (Z/\gamma)^*\to\ConeCone$
and $q\bar{q}'\to W^{\pm*}\to\ConeNtwo$.  
The chargino and neutralino decay properties depend on the Minimal Supersymmetric Standard Model (MSSM) parameters $M_1$ and $M_2$ (the gaugino
masses), $\tan{\beta}$ (the ratio of the vacuum expectation values of the
two Higgs doublets), and  $\mu$ (the higgsino-mixing mass term).  
In this paper we study two pMSSM model implementations with large $\tan{\beta}$ and where the only light slepton is the stau partner of the right-handed tau ($\tilde{\tau}_{\rm R}$). 
More details of the considered models are given in section \ref{sec:signalGrids}.

\begin{figure}\centering
	\raisebox{0.15\textwidth}{(a)}\includegraphics[width=0.35\textwidth]{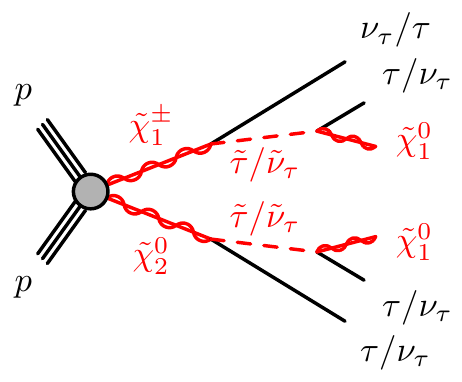}
	\raisebox{0.15\textwidth}{(b)}\includegraphics[width=0.35\textwidth]{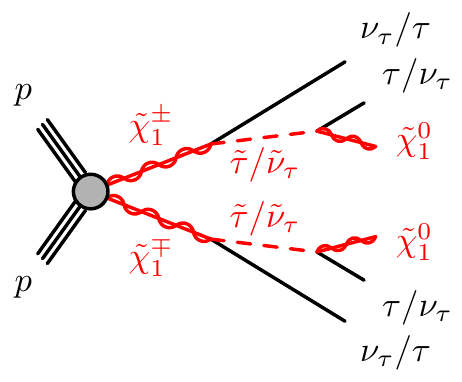}\hfil
	\raisebox{0.15\textwidth}{(c)}\includegraphics[width=0.35\textwidth]{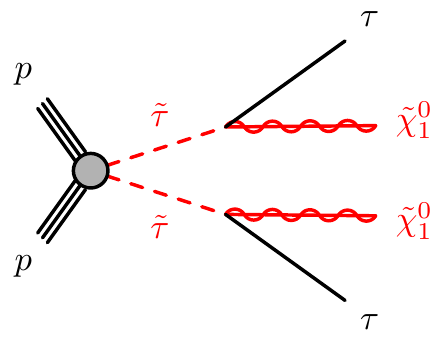}\hfil
	\caption{\label{diagrams}
		Representative diagrams for the electroweak production processes of supersymmetric particles considered in this work:
		(a) $\susy{\chi}^{\pm}_{1}\susy{\chi}^{0}_{2}$, (b) $\susy{\chi}^{\pm}_{1}\susy{\chi}^{\mp}_{1}$, and (c) $\tilde{\tau}\tilde{\tau}$ production.}
\end{figure}

``Simplified models''~\cite{simplified,Alves:2011wf} characterised by \chinoonepm\ninotwo\ and \chinoonepm\chinoonemp\ production, where the charginos and neutralinos decay with 100\% 
branching fraction to final states with taus, are also considered.
In both simplified models the lightest neutralino is the LSP, and the only light slepton is the stau partner of the left-handed tau ($\tilde{\tau}_{\rm L}$).
If the stau and the superpartner of the tau neutrino (the tau sneutrino, labelled as $\tilde{\nu}_\tau$) are lighter than the \chargino{1} and \neutralino{2}, the following decay processes can occur:
$\tilde{\chi}_{2}^{0} \rightarrow \tilde{\tau}_{\rm L}\tau \rightarrow  \tau\tau\tilde{\chi}_{1}^{0}$,
and $\tilde{\chi}_{1}^{\pm} \rightarrow  \tilde{\tau}_{\rm L}\nu (\tilde{\nu}_\tau \tau) \rightarrow \tau\nu\tilde{\chi}_{1}^{0}$.
Figures~\ref{diagrams}(a) and~\ref{diagrams}(b) show diagrams of associated $\susy{\chi}^{\pm}_{1}\susy{\chi}^{0}_{2}$ and $\susy{\chi}^{\pm}_{1}\susy{\chi}^{\mp}_{1}$ production. 

If charginos and next-to-lightest neutralinos are too heavy to be produced at the LHC, direct production of stau pairs~\cite{slepton_production} 
might become the dominant electroweak production process in the pMSSM. The production of stau pairs is studied in this paper, and the relevant
process is depicted in figure~\ref{diagrams}(c). 

The studied final state contains at least two taus with opposite electric charge (opposite-sign, OS), low jet activity and large missing 
transverse momentum due to the escaping neutrinos and LSPs. Only final states containing hadronically decaying taus are 
considered in this search. 

\section{The ATLAS detector}
\label{sec:atlas}

The ATLAS detector~\cite{Aad:2008zzm} is a multi-purpose particle physics detector with 
forward-backward symmetric cylindrical geometry, and nearly $4\pi$
coverage in solid angle.\footnote{%
	ATLAS uses a right-handed coordinate system with its origin at the nominal 
	interaction point (IP) in the centre of the detector, and the $z$-axis along the 
	beam line. The $x$-axis points from the IP to the centre of the LHC ring, 
	and the $y$-axis points upwards. Cylindrical coordinates $(r,\phi)$ are used 
	in the transverse plane, $\phi$ being the azimuthal angle around the $z$-axis. 
	Observables labelled ``transverse'' refer to the projection into the $x$--$y$ plane. 
	The pseudorapidity is defined in terms of the polar angle $\theta$ by $\eta=-\ln\tan(\theta/2)$.} 
It features an inner tracking detector (ID) surrounded by a 2 T superconducting solenoid, 
electromagnetic and hadronic calorimeters, and a muon spectrometer (MS).  The ID covers the pseudorapidity region  $|\eta| <2.5$ and consists of a 
silicon pixel detector, a silicon microstrip detector (SCT), and a transition
radiation tracker (TRT). The calorimeters are composed of high-granularity liquid-argon (LAr) electromagnetic calorimeters with lead, copper, 
or tungsten absorbers (in the pseudorapidity region $|\eta| < 3.2$) and an iron--scintillator hadronic calorimeter (over
$|\eta|<1.7$). 
The end-cap and forward regions, spanning
$1.5<|\eta|<4.9$, are instrumented with LAr calorimeters
for both the electromagnetic and hadronic measurements. The MS surrounds the calorimeters and consists 
of three large superconducting air-core toroid magnets, each with eight coils, a system of precision tracking chambers ($|\eta|<2.7$), and detectors for triggering ($|\eta|<2.4$).
Events are selected by a three-level trigger system.

\section{Data sample}
\label{sec:data}

The analysed dataset, after the application of beam, detector and data quality
requirements,  corresponds to an integrated luminosity of $20.3 \pm 0.6~\ifb$.
 The luminosity is measured using techniques similar to those described in ref.~\cite{lumi2012} with a preliminary calibration of the luminosity scale 
 derived from beam-overlap scans performed in November 2012. 
The average number of inelastic interactions per bunch crossing (pile-up) varied from 5.9 to 36.5.

The events used in this analysis are recorded using a di-tau trigger, which requires identification of two
hadronically decaying tau candidates with transverse momenta (\pt)
exceeding a set of thresholds, similar to those described in ref. \cite{tautrigger}.
Trigger efficiency measurements using a sample of $Z\rightarrow \tau\tau$ events where one tau decays hadronically and the other leptonically 
into a muon and two neutrinos, show that the di-tau trigger reaches constant efficiency ($\sim$65\%) when the leading tau has 
 $\pt > 40$~\GeV~and the next-to-leading tau has $\pt > 25$~\GeV.

\section{Monte Carlo simulation}

Monte Carlo (MC) simulated event samples are used to estimate the SUSY signal yields
and to aid in evaluating the SM backgrounds.
MC samples are processed through a detailed detector simulation \cite{atlassimulation} based on {\tt GEANT4} \cite{geant4} and reconstructed using the
same algorithms as the data. The effect of multiple proton--proton collisions in the same or nearby bunch crossings is also taken
into account.

\subsection{Standard Model processes}

The main sources of SM background to final states with at least two
hadronically decaying taus are multi-jet, $W$+jets and diboson events. They are 
estimated with methods using simulation samples and data as described in section \ref{sec:backgrounds}.
MC samples are used to estimate the SM background contributions from processes
leading to at least one tau from prompt boson decays in the final state, such as diboson production ($WW$, $WZ$, $ZZ$), processes including 
a top quark pair or single top quark (in association with jets or $W$/$Z$ bosons), and $Z$ boson production in association with jets. 
Production of the SM Higgs boson with a mass of 125 GeV is also considered.

The diboson samples are generated with {\tt SHERPA} v1.4.1~\cite{sherpa}, with 
additional gluon--gluon contributions simulated with \texttt{gg2WW} v3.1.2~\cite{gg2WW}. 
The production of top quark pairs is also simulated with the {\tt SHERPA} v1.4.1 generator.
Samples of \ttbar+$V$ ($V=W,Z$) are generated with the leading-order (LO) generator {\tt MadGraph 5} v1.3.33 
\cite{madgraph} interfaced to {\tt PYTHIA} v8.165~\cite{pythia,PYTHIA8}. Single-top production is simulated with {\tt MC@NLO} v4.06 
 ($Wt$- and $s$-channel) \cite{mcatnlo1,mcatnlo2,mcatnlo3} and {\tt AcerMC} v3.8 ($t$-channel) \cite{Kersevan2013919}. 
In all samples the top quark mass is set to $172.5$~GeV.
Events with $Z/\gamma^*\to\ell\ell$ and $W\to\ell\nu$ produced with accompanying jets (including light and heavy flavours) are generated with 
\alpgen\ v2.14~\cite{alpgen} interfaced to {\tt PYTHIA} 6. 
The gluon fusion and vector-boson fusion production modes of the SM Higgs
are simulated with \powheg-\texttt{BOX} v1.0~\cite{powheg}, and the
associated production ($WH$ and $ZH$) with {\tt PYTHIA} v8.165.

The simulation parameters are tuned to describe the soft component of
the hadronic final state~\cite{mc11tune, Skands:2010ak}.
The next-to-leading-order (NLO) CT10 \cite{CT10} parton distribution function (PDF) set is used for {\tt SHERPA} and {\tt MC@NLO}. 
The CTEQ6L1 \cite{CTEQ6} set is used for {\tt MadGraph}, {\tt AcerMC}, {\tt PYTHIA}, and  {\tt ALPGEN}.

All SM background production cross sections are normalised to the results of higher-order calculations when 
available. The inclusive $W$ and $Z$ production cross sections are calculated to next-to-next-to-leading order (NNLO) in the strong coupling
constant with {\tt DYNNLO}  \cite{Catani:2009sm} using the MSTW2008NNLO PDF set \cite{MSTW}.
The \ttbar\ cross section is normalised to a NNLO calculation including resummation of next-to-next-to-leading logarithmic (NNLL) 
soft gluon terms obtained with \texttt{Top++}
v2.0~\cite{Czakon:2011xx}. The diboson production cross section is normalised to NLO
using {\tt MCFM} v6.2 \cite{Campbell:1999ah,Campbell:2011bn}.
The production of \ttbar\ in association with $W/Z$ is normalised to the NLO cross section \cite{Campbell:2012dh,Garzelli:2012bn}.

\subsection{SUSY processes}
\label{sec:signalGrids}

Simulated signal samples are generated with \texttt{Herwig++} v2.5.2~\cite{Bahr:2008pv} and the CTEQ6L1 PDF set.
Signal production cross sections are calculated to NLO using \texttt{PROSPINO2}~\cite{prospinoCNSL}. 
They are in agreement with the NLO calculations matched to resummation at the next-to-leading-logarithmic accuracy (NLO+NLL) within $\sim$2\% \cite{Fuks:2012qx,Fuks:2013vua,Fuks:2013lya}.

The results of this search are interpreted in the context of two pMSSM models with the following specifications. 
The masses of squarks, gluinos and sleptons other than the stau partner of the right-handed taus are set to 3 TeV, and $\tan \beta$ is set to 50.
In the first (second) pMSSM model, $M_1$ is set to 50 (75) GeV and $M_2$ and $\mu$ are varied between 100 and 500 (600) GeV. 
The mass of the lighter stau, $\tilde{\tau}_1$ = $\tilde{\tau}_{\rm R}$, is set to 95 GeV in the first pMSSM model, whereas in the second pMSSM model it is set halfway between those of the $\tilde{\chi}_{2}^{0}$ and the 
\ninoone. 
In the first pMSSM scenario (with fixed stau mass), the dominant processes are the associated production of charginos and neutralinos
($\tilde{\chi}_{1}^{\pm}\tilde{\chi}_{2}^{0}$), or pair production of
charginos ($\tilde{\chi}_{1}^{\pm}\tilde{\chi}_{1}^{\mp}$) or staus ($\tilde{\tau}\tilde{\tau}$), depending on the values of $M_2$ and $\mu$.
The cross section of direct stau production is 163 fb over the whole set of models, while the production cross sections of chargino--neutralino and chargino--chargino
vary from 5~$\cdot~10^{-3}$ to 40 pb and from 0.01 to 16 pb, respectively.
In the second pMSSM scenario (with variable stau mass), the dominant processes are 
$\tilde{\chi}_{1}^{\pm}\tilde{\chi}_{2}^{0}$ and $\tilde{\chi}_{1}^{\pm}\tilde{\chi}_{1}^{\mp}$ production.
The cross section of direct stau production varies from 0.4 to 42 fb, 
while the production cross sections of chargino--neutralino and chargino--chargino
vary from 5~$\cdot~10^{-4}$ to 1.2 pb and 8~$\cdot~10^{-4}$ to 0.9 pb, respectively.

Two simplified models characterised by $\tilde{\chi}_{1}^{\pm}\tilde{\chi}_{2}^{0}$
and $\tilde{\chi}_{1}^{\pm}\tilde{\chi}_{1}^{\mp}$ production are also considered. In these models, all sparticles other 
than \chinoonepm, \ninotwo, \ninoone, $\tilde{\tau}_{\rm L}$ and $\tilde{\nu}_\tau$ are assumed to 
be heavy (masses of order of 2 TeV). 
The neutralinos and charginos decay via intermediate staus and tau sneutrinos. The stau 
and tau sneutrino are assumed to be mass-degenerate, which happens to be often the case in pMSSM scenarios with large mass splitting between
\chinoonepm\ and \ninoone. The mass of the $\tilde{\tau}_{\rm L}$ state is set to be halfway between those of the \chinoonepm\ and the \ninoone.
Furthermore, \chinoonepm\ and \ninotwo\ are assumed to be pure wino and mass-degenerate, while the  \ninoone\ is purely bino.
The \chinoonepm\ (\ninotwo) mass is varied between 100 and 500 GeV, and the \ninoone\ mass is varied between zero and 350 GeV.
The cross section for electroweak production of supersymmetric
particles ranges from 0.01 to 2 pb in the considered models.

Direct stau production is studied in the context of the pMSSM model described in ref.~\cite{AbdusSalam2011}.
The masses of all charginos and neutralinos apart from the $\neutralino{1}$ are set to $2.5\TeV$. The model contains $\tilde{\tau}_{\rm R}$
and $\tilde{\tau}_{\rm L}$, but no tau sneutrinos. 
The stau mixing is set such that $\tilde{\tau}_1 = \tilde{\tau}_{\rm R}$ and $\tilde{\tau}_2 = \tilde{\tau}_{\rm L}$.
The stau masses are generated in the range from 90 to 300 GeV and the mass of the bino-like $\neutralino{1}$ is varied by scanning the 
gaugino mass parameter $M_1$ in the range from zero to 200 GeV.
The cross section for direct stau pair production in this scenario
decreases from 176 to 1.4 fb for $\tilde{\tau}_{\rm L}$, and from 70 to 0.6 fb
for $\tilde{\tau}_{\rm R}$ as the stau mass increases from 90 to 300 GeV.

Three reference points are used throughout this paper to illustrate the typical features of the
 SUSY models to which this analysis is sensitive:

\begin{itemize}
 \item Ref. point 1: simplified model for chargino--neutralino production with mass of \chinoonepm\ (\ninotwo) equal to 250 GeV, and mass of
 \ninoone\ equal to 100 GeV;
 \item Ref. point 2: simplified model for chargino--chargino production with mass of \chinoonepm\ equal to 250 GeV, and mass of
 \ninoone\ equal to 50 GeV;
 \item Ref. point 3: direct stau production with mass of the  $\tilde{\tau}_{\rm R}$ ($\tilde{\tau}_{\rm L}$) equal to 127 (129) GeV, and massless \ninoone. 
\end{itemize}

\section{Event reconstruction}
\label{sec:object} 

\newcommand{\pTvec}{\mathbf{p}_\mathrm{T}}
\newcommand{\qTvec}{\mathbf{q}_\mathrm{T}}
\newcommand{\qT}{q_\mathrm{T}}
\newcommand{\mT}{m_\mathrm{T}}
\newcommand{\pTell}[1]{\mathbf{p}_\mathrm{T}^{\ell#1}}
\newcommand{\pTmiss}{\mathbf{p}_\mathrm{T}^\mathrm{miss}}

Events with at least one reconstructed primary vertex are selected. A primary vertex must have 
at least five associated charged-particle tracks with $\pt>$~400~MeV and be consistent with the beam spot envelope. 
If there are multiple primary vertices in an event, the one with the largest $\sum\pt^2$ of the associated
tracks is chosen.

Jets are reconstructed from three-dimensional calorimeter energy clusters using the anti-$k_t$ algorithm~\cite{anti-kt,anti-kt2} 
with a radius parameter of 0.4. Jet energies are corrected for detector inhomogeneities, 
the non-compensating nature of the calorimeter, and the impact of pile-up, using factors derived from test beam, cosmic ray, 
and {\it pp} collision data, and from a detailed {\tt GEANT4} detector simulation \cite{Aad:2011he}.
The impact of pile-up is accounted for by using a technique, based on jet areas, that provides an 
event-by-event and jet-by-jet correction \cite{jetarea}.
Events containing jets that are likely to have arisen from detector noise or cosmic rays are removed. 
For this analysis jets are required to have $\pt>20\GeV$ and $\vert\eta\vert < 4.5$.

Electron candidates are reconstructed by matching clusters in the electromagnetic calorimeter 
with charged particle tracks in the inner detector. Electrons are required to have $\pt>10\GeV$, $|\eta|<2.47$, and to satisfy the 
``medium'' shower-shape and track-selection criteria defined in ref.~\cite{egammaperf}, updated for the 2012 operating conditions.
Muon candidates are identified by matching an extrapolated inner detector track and one or more track segments in the muon 
spectrometer \cite{muon2012}. Muons are required to have $\pt > 10\GeV$ and $\vert\eta\vert < 2.4$.  
Events with muons compatible with cosmic rays are rejected.

The reconstruction of hadronically decaying taus is based on the information from tracking in the ID and three-dimensional clusters 
in the electromagnetic and hadronic calorimeters.
The tau reconstruction algorithm is seeded by jets reconstructed as described above but with $p_{\rm T}>10\GeV$ and $|\eta|<2.47$. Tracks are subsequently associated with the
tau jet within a cone of size $\Delta R =\sqrt{(\Delta \eta)^2+(\Delta \phi)^2}=0.2$ around the axis of the tau cluster. 
The reconstructed energies of the hadronically decaying tau candidates are corrected to the tau energy scale, which is calibrated independently of 
the jet energy scale, by a MC-based procedure \cite{atlas-tau-energyscale}. Tau neutrinos from the tau lepton decay are not taken into account in 
the reconstruction and calibration of the tau energy and momentum.
Since taus decay mostly to either one or three charged 
pions, together with a neutrino and often additional neutral pions, tau candidates are required to have one or three associated charged particle tracks (prongs)
 and the total electric charge of those tracks must be $\pm 1$ times the electron charge.
To improve the discrimination between hadronically decaying taus and jets,
electrons, or muons, multivariate algorithms are used
\cite{tau-reco+id}. The tau identification algorithm used in this analysis is based on the Boosted Decision Tree (BDT) method.
The BDT algorithms use as input various track and cluster variables for particle discrimination. 
A ``jet BDT'' is used to discriminate taus from jets, and an ``electron BDT'' to discriminate between electrons and taus. 
Based on the jet BDT result, three tau identification criteria corresponding to ``loose'', ``medium'', and ``tight'' quality 
can be defined.
For 1-prong (3-prong) taus the signal efficiencies are 70\%, 60\% and 40\% (65\%, 55\% and 35\%) for the 
loose, medium and tight working points, respectively. 
Background rejection factors ranging from 10 to 40 for signal efficiencies of 70\% are achieved, increasing to 500 for 35\% signal efficiency.
In the following, tau candidates are required to pass the ``medium'' identification criteria for jet discrimination, while for the final signal region 
selections both the ``tight'' and ``medium'' criteria are used.
For electron discrimination, the ``loose'' quality selection is applied to  1-prong taus only.
This requirement has about 95\% efficiency, and a rejection factor from 10 to 50 depending on the $\eta$ range.
In addition, a dedicated muon veto is applied to remove tau candidates generated by muons associated with anomalous energy deposits in the
calorimeter. 
The resulting signal efficiency is better than 96\%, with a reduction of muons misidentified as taus of around 40\%.
Tau candidates are required to have $\pt > 20\GeV$ and $\vert\eta\vert < 2.47$.

The measurement of the missing transverse momentum two-vector, $\pTmiss$, and its magnitude, $\met$, is based on the vectorial sum of the $\bold{\pT}$ of 
reconstructed objects (jets, taus, electrons, photons, muons) as well as
calorimeter energy clusters (with $\left|\eta\right| < 4.9$) not associated with reconstructed objects \cite{Aad:2012re}.
Since different requirements on the tau identification (loose, medium, and tight jet BDT quality requirements) are used throughout this analysis, taus are calibrated at the jet energy scale for the calculation of $\pTmiss$.

The possible double counting of reconstructed objects is resolved in the following order.
If two electron candidates are found within a distance $\Delta R=$ 0.05, the electron candidate with lower momentum is discarded.
Jet candidates are removed if they lie within a distance $\Delta R =$ 0.2 of a tau or an electron. 
Any tau candidate lying within a distance $\Delta R=$ 0.2 of any remaining electron or muon is discarded.
Any muon candidate within $\Delta R=$ 0.4 of a jet is removed.
The remaining electrons are rejected if they lie within 0.2 $< \Delta R < 0.4$ of a jet. 
Electron and muon candidates within a distance $\Delta R =$ 0.1, or muon candidates within a distance $\Delta R =$ 0.05, are rejected.
To remove low-mass resonances, electron or muon pairs with opposite electrical charge are rejected 
if their invariant mass is less than 12 GeV. 

Jets are further classified as jets containing a $b$-quark ($b$-jets), light-parton jets, and forward jets.
A $b$-tagging algorithm~\cite{btag7tev}, which exploits the long lifetime of $b$-hadrons, is used to identify jets containing a $b$-quark. 
The mean nominal $b$-tagging efficiency, determined from \ttbar\ MC events, is 80\%, with a misidentification (mis-tag) rate for 
light-quark/gluon jets of less than 1\%. 
Correction factors are applied as functions of the \pt\ and $\eta$ of the jets to all MC samples to correct for small differences in the 
$b$-tagging performance observed between data and simulation.
Jets in the central region, satisfying $\pt>20\GeV$, $|\eta|<2.4$, and the $b$-tagging algorithm, are defined as
$b$-jets, or \verb|B20|.
Central light jets are required to have $|\eta|<2.4$, and not be identified as $b$-jets.
To remove jets that originated from pile-up collisions, a central light jet with $\pT<50\GeV$ 
must have at least one track with $\pt>$~400~MeV associated with it and with the primary vertex of the event.
 Depending on their transverse momentum, central light jets are referred to as \verb|L30| (\verb|L50|) if they have 
$\pt>$ 30 (50)$\GeV$. 
Forward jets, or \verb|F30|, must satisfy $\pt>30\GeV$ and $2.4<|\eta|<4.5$.

For the background estimation and validation described in section \ref{sec:backgrounds}, events containing isolated electrons or 
muons are selected. Isolated electrons and muons are defined as follows. The summed scalar $\pT$ of tracks above $400\MeV$ within a 
cone of size $\Delta R=0.3$ around each electron or muon candidate (excluding the candidate itself) and associated with the primary 
vertex is required to be less than 16\% of the electron or muon $\pT$. 
The distance of closest approach in the transverse plane of an electron or muon candidate to the event primary vertex must be within five (for electron
candidates) or three (for muon candidates) standard deviations from its measurement in the transverse plane.
For isolated electrons, the sum of transverse energies of the surrounding calorimeter energy clusters within $\Delta R =0.3$ of each electron candidate, 
corrected for the deposition of energy from pile-up interactions, is required to be less than 18\% of the electron \pt.
The longitudinal impact parameter of an electron, $z_0$, must satisfy $|z_0\sin\theta|<0.4$\,mm.
Muon candidates are required to satisfy $|z_0\sin\theta|<0.1\,\mathrm{mm}$.
Furthermore, isolated electrons must satisfy the ``tight'' criteria~\cite{egammaperf, egammaperf1} 
placed on the ratio of calorimetric energy to track momentum, and the number of high-threshold hits in the TRT\@.

The simulation is corrected for differences in the efficiency of the tau identification and trigger algorithms between data and MC simulation. 
For hadronically decaying taus coming from prompt boson decays, the corrections are calculated with a ``tag-and-probe'' method in a sample of 
$Z \rightarrow \tau \tau$ events where one tau decays hadronically and the other 
leptonically into a muon and two neutrinos  \cite{tau-reco+id}. For misidentified taus from electrons a sample of $Z \rightarrow ee$ events 
is used, while the mis-identification rate of taus from light jets is measured in a sample of $Z \rightarrow \mu
\mu$ events with associated jets. 
The efficiencies for electrons and muons to satisfy the reconstruction, identification and isolation criteria are measured in samples of $Z$ and $J/\psi$
leptonic decays, and corrections are applied to the simulated samples to reproduce the efficiencies in data. 

\section{Event selection}
\label{sec:SR}

Events are required to have at least two candidate taus, and at least one must satisfy the tight jet BDT quality requirement.
At least one of the selected tau pairs must contain taus with opposite electrical charge. Two of the reconstructed taus must have fired the di-tau trigger,
and satisfy the $\pt$ requirements to be in the region where the trigger efficiency is constant, i.e. the more (less) energetic tau must have 
 $\pt >$ 40 (25) GeV. 
The di-tau invariant mass of any opposite-sign (OS) pair must be larger than $12\GeV$ to remove taus from low-mass resonances.
This requirement has negligible effect on the signal efficiency.
Events with additional light leptons (defined as candidate electrons or muons after resolving the overlap between the various reconstructed 
objects as described in section \ref{sec:object}) are vetoed to allow
for a statistical combination with other ATLAS analyses~\cite{2Lep-2012,3Lep-2012}.

To enhance the sensitivity to the SUSY signal and suppress SM backgrounds, additional requirements are applied that define the so-called signal
regions (SR). At tree level, no jet is present in the SUSY processes of interest; however, jets can be generated from initial-state 
radiation (ISR).
Vetoing events containing jets suppresses background contributions involving top quarks but may also reduce the signal efficiency due to ISR.
To maximise the signal-to-background ratio, three event-based jet vetoes are used depending on the SR: a ``$b$-jet veto'', where only events with 
no $b$-jets, i.e. $N$(\verb|B20|) = 0, are accepted; a ``jet veto'', where only events with
$N$(\verb|B20|)+$N$(\verb|L30|)+$N$(\verb|F30|) = 0 are selected; and a 
``looser jet-veto'', where $N$(\verb|B20|)+$N$(\verb|L50|)+$N$(\verb|F30|) = 0 is required.
The rejection factors of the $b$-jet veto, jet veto, and looser jet-veto measured on MC samples of top quark backgrounds (\ensuremath{t\bar{t}}, 
single top and \ensuremath{t\bar{t}}+$V$) are about 6, 50, and 20, respectively. 

To reject backgrounds with a $Z$ boson, events where at least one of
the oppositely charged tau pairs has a reconstructed invariant mass within 10 GeV of the visible 
$Z$ boson mass (81 GeV) are vetoed. 
The visible $Z$ boson mass is obtained from the mean value of a gaussian fit of the reconstructed invariant mass distribution of OS tau pairs in a MC sample of 
$Z( \rightarrow \tau \tau)$+jets events. This requirement is referred to as the ``$Z$-veto''.
The $Z$-veto reduces the contribution from background processes containing $Z$ decays by a factor of three.
An additional requirement on the angular separation between the leading  and  next-to-leading tau, $\Delta R(\tau,\tau)<3$,
is effective in discriminating against back-to-back events such as multi-jet production or $Z$ decays.

To further improve the signal to background ratio, additional requirements on a selection of sensitive kinematic variables
are applied depending on the SR. 
The ``stransverse'' mass \mttwo~\cite{Lester:1999tx,Barr:2003rg} is defined as:
\[
\mttwo = \min_{\qTvec}\left[\max\left(\mT(\pTvec^{\tau1},\qTvec),\mT(\pTvec^{\tau2},\pTmiss-\qTvec)\right)\right],
\]
where $\pTvec^{\tau1}$ and $\pTvec^{\tau2}$ are the transverse momenta of the two taus,
and $\qTvec$ is a transverse vector that minimises the larger of the
two transverse masses $\mT$.
The latter is defined by
\[
\mT(\pTvec,\qTvec) = \sqrt{2(\pT\qT-\pTvec\cdot\qTvec)}.
\]
In events where more than two taus are selected,  \mttwo\ is computed among all possible tau pairs and the combination leading to the largest value is chosen.
For \ttbar\ and $WW$ events, in which two on-shell $W$ bosons decay leptonically and $\pTmiss$
is the sum of the transverse momenta of the two neutrinos, the \mttwo\ distribution has an upper end-point at the $W$ mass.
For large mass differences between the next-to-lightest neutralinos, the charginos, or the staus and the lightest neutralino, the \mttwo\
distribution for signal events extends significantly beyond the distributions of the \ttbar\ and $WW$ events.

Two additional kinematic variables, sensitive to the additional missing transverse momentum due to the LSPs, are used to discriminate SUSY from 
SM events. These are the effective mass $m_{\rm eff} = \met +
\pT^{\tau1} + \pT^{\tau2}$, defined as the scalar sum of the missing transverse energy 
and the transverse momenta of the two leading taus, and $m_{{\rm
    T}\tau1}+m_{{\rm T}\tau2}$, where $m_{{\rm T}\tau1}$ ($m_{{\rm
    T}\tau2}$) is the transverse mass computed from the transverse
momentum of the leading (next-to-leading) tau and $\pTmiss$. The
correlation among the above defined kinematical variables varies between 20\% and 90\% according to the signal model.

Four SRs are defined in this analysis, targeting various SUSY production processes.\footnote{In the SR definitions, the
following mnemonic naming conventions are used: ``C1'' stands for $\tilde{\chi}^{\pm}_{1}$, ``N2'' for $\tilde{\chi}^{0}_{2}$,
and ``DS'' for direct stau production.} 
The requirements for each SR are summarised in table~\ref{tab:SR-def}. 
At least two OS taus are required in all SRs except for SR-C1C1, where exactly  two OS taus are required.
The first two SRs, SR-C1N2 and SR-C1C1, target $\tilde{\chi}_{1}^{\pm}$$\tilde{\chi}_{2}^{0}$ and $\tilde{\chi}_{1}^{\pm}$$\tilde{\chi}_{1}^{\mp}$ production, respectively. 
SR-C1N2 is based on moderate $\met$ and large \mttwo\ requirements. 
In SR-C1C1, events with moderate \mttwo\ and large $m_{{\rm T}\tau1}+m_{{\rm T}\tau2}$ are selected. 
The last two SRs, SR-DS-highMass and SR-DS-lowMass, are designed to cover direct stau production and are optimised for different ranges of the stau 
mass. The four SRs are not mutually exclusive.

The signal acceptance for events passing all analysis requirements is of order a few percent 
for SUSY models to which this analysis is sensitive in all SRs. The trigger efficiency varies
between 55\% and 70\% according to the signal region and the SUSY
model considered, while the total reconstruction efficiency is about 15\%. 

\begin{table}[h]
\caption{\label{tab:SR-def} Signal region definitions.}
\begin{center}
\begin{tabular}{|c|c|c|c|}
\hline
SR-C1N2 		  &  SR-C1C1         & SR-DS-highMass         & SR-DS-lowMass  \\
\hline
\hline
  $\ge2$ OS taus	  &  2 OS taus         & $\ge2$ OS taus   & $\ge2$ OS taus \\
  $b$-jet veto  	  &  jet veto	           & looser jet-veto	      & looser jet-veto \\
  $Z$-veto		  &  $Z$-veto	       & $Z$-veto	              & $Z$-veto\\
			  &		       & $\Delta R(\tau,\tau)<3$& $\Delta R(\tau,\tau)<3$ \\
  $\met >40$~GeV	  &  \mttwo~$>30$~GeV  & \mttwo~$>60$~GeV        & \mttwo~$>30$~GeV \\
 \mttwo~$>100$~GeV & $m_{{\rm T}\tau1}+m_{{\rm T}\tau2}>250$~GeV 	  & $m_{\rm eff}>230$~GeV      & $m_{\rm eff}>260$~GeV \\
\hline
\end{tabular}

\end{center}
\end{table}

\section{Standard Model background estimation}\label{sec:backgrounds}

The main SM processes contributing to the selected final states are
multi-jet, $W$+jets and diboson production.
Decays of the SM Higgs boson, assuming a mass of 125 GeV, into a $\tau\tau$ final state have negligible contribution in all SRs (less than 0.1\%).
Background events may contain a combination of `real' taus, defined as
correctly identified tau leptons, or `fake' taus, which can originate from a 
misidentified light-flavour quark or gluon jet, an electron or a muon. 

In multi-jet events all tau candidates are misidentified jets. Due to the large cross section and the poor MC modelling of the tau
mis-identification rate from jets, the multi-jet contribution in the SRs is estimated from data, as described in section
\ref{sec:BG_QCD}.
The contribution arising from heavy-flavour multi-jet events containing
a real tau lepton from the heavy-flavour quark decay is included in the
multi-jet estimate. 
The contribution of $W$+jets events, which contain one real tau from
the $W$ decay and one or more misidentified jets, is estimated from MC
simulation, and normalised to data in a dedicated control region, as described in section \ref{sec:BG_W}.

Diboson production contributes mainly with events containing real tau leptons coming from
$WW$ and $ZZ$ decaying into a $\tau\tau\nu\nu$ final state.
Additional SM backgrounds arise from $Z$+jets production, or events which contain a top quark or top quark pair in association with jets or
additional $W$ or $Z$ bosons (collectively referred to as `top'  background in the following). 
The contribution from real taus exceeds 90\% in $Z$+jets and diboson production, and ranges from 45\% to 75\% in backgrounds containing
top quarks. The contribution of fake taus from heavy-flavour decays in jets is negligible.
To estimate the irreducible background, which includes diboson, $Z$+jets and top quark events, only MC simulated samples are used, as
described in section \ref{sec:BG_irre}. The available MC sample sizes in the SRs are limited for $Z$+jets and top backgrounds.
To improve the statistical precision, the prediction in the SRs for these processes is extrapolated from regions with large MC statistics. 

Finally, for each signal region, a simultaneous fit based on the profile likelihood method \cite{statforumlimitsl} is performed to normalise
the multi-jet and $W$+jets background estimates, as described in section \ref{sec:BG_fit}.
Details of the sources of systematic uncertainty on the background estimates described in this section are given in section \ref{sec:Systematics}.

\subsection{Multi-jet background estimation}
\label{sec:BG_QCD}

One of the dominant backgrounds in the SRs originates from jets misidentified as taus in multi-jet production (from 
13\% to 30\% of the total background, depending on the SR). 
This contribution is estimated from data using the ``ABCD'' method. 
Four exclusive regions, labelled as A, B, C (the ``control regions'') and D (the SR), are defined in a two-dimensional plane 
as a function of two uncorrelated discriminating variables.
In this case, the ratio of the numbers of events in the control regions (CR) A and B equals that of 
SR D to control region C: the number of events in the SR D, $N_{\rm D}$, can therefore be calculated from that in control region A, $N_{\rm A}$,
multiplied by the transfer factor $\rm{T}=N_{\rm C}/N_{\rm B}$. 

The tau identification criterion (tau-id) based on the jet BDT quality requirement and a kinematic variable chosen depending on the SR are used
 as the two discriminating variables to define the regions A, B, C and D.
The following kinematic variables are used: \mttwo\ for SR-C1N2, 
$m_{{\rm T}\tau1}+m_{{\rm T}\tau2}$  for SR-C1C1, 
and $m_{\rm eff}$ for SR-DS-highMass and SR-DS-lowMass. 

The control region A and the signal region D are defined in the same way except that in the control region A
all candidate taus must satisfy the ``loose'' but fail the ``tight'' jet BDT requirement (tight tau event veto).
The same requirement on the tau-id as in control region A (signal region D) is applied in control region B (control region C). 
In control regions B and C, less stringent requirements on the kinematic variables defined above are applied. 
The definitions of the control regions are summarised in table \ref{tab:QCD-CR}.

Furthermore, two validation regions E and F are defined. The validation region E (F)
has the same definition as the control region A (signal region D) 
except for intermediate requirements on the kinematic variable, as listed in table \ref{tab:QCD-VR}.
The validation regions are used to verify the extrapolation of the ABCD estimation to the SR, and to estimate the systematic 
uncertainty from the residual correlation between the tau-id and the kinematic variable.
The regions A--F are drawn schematically in figure \ref{fig:abcd}.

\begin{table}
\caption{The multi-jet control region definitions. Only those requirements that are different in the CRs with respect to the SRs are listed.}
\smallskip
\centering
\footnotesize
\begin{tabular}{c|c|c|c}\hline
Regions &A&B&C\\ 
\hline
 CR-        & $\mttwo >$~100 GeV     &  $\mttwo <$~40 GeV           & $\mttwo <$~40 GeV           \\
C1N2        &  at least 2 loose taus &  at least 2 loose taus       &  at least 1 medium tau        \\ 
            &  tight tau veto        &  tight tau veto              & at least 1 tight tau         \\\hline 
CR-        & $m_{{\rm T}\tau1}+m_{{\rm T}\tau2} > $ 250 GeV & 80 $< m_{{\rm T}\tau1}+m_{{\rm T}\tau2} <$ 150 GeV & 80  $< m_{{\rm T}\tau1}+m_{{\rm T}\tau2} <$ 150 GeV \\
C1C1     & at least 2 loose taus  & at least 2 loose taus        & at least 1 medium tau          \\
             &  tight tau veto        &  tight tau veto              &  at least 1 tight tau         \\\hline  
CR-DS-      & $m_{\rm eff} >$ 230 GeV  & 130 GeV $< m_{\rm eff}<$ 150 GeV & 130 GeV $< m_{\rm eff} <$ 150 GeV    \\
highMass    &  at least 2 loose taus & at least 2 loose taus        & at least 1 medium tau        \\
            &  tight tau veto        &  tight tau veto              & at least 1 tight tau        \\ \hline		
CR-DS-      & $m_{\rm eff} >$ 260 GeV  & 100 GeV $< m_{\rm eff}<$ 150 GeV & 100 GeV $< m_{\rm eff} <$ 150 GeV     \\ 
lowMass     &  at least 2 loose taus & at least 2 loose taus        & at least 1 medium tau        \\ 
            &  tight tau veto        &  tight tau veto              & at least 1 tight tau        \\ \hline
\end{tabular}

\label{tab:QCD-CR}
\end{table}
\begin{table}
\caption{The requirement on the kinematic variables used to define the validation regions E and F. Only those requirements that are different with 
respect to the A, B, and C CRs and the SRs are listed.}
\smallskip
\centering
\vspace{0.1cm}
\begin{tabular}{c|c}\hline
Regions          & E/F \\ \hline
 VR-C1N2         & 40 GeV $< \mttwo <$ 100 GeV                     \\
 VR-C1C1         & 150 GeV $< m_{{\rm T}\tau1}+m_{{\rm T}\tau2} < $ 250 GeV      \\ 
 VR-DS-highMass   & 150 GeV $<m_{\rm eff} < 230$ GeV       \\
 VR-DS-lowMass    & 150 GeV $<m_{\rm eff} < 260$ GeV        \\ \hline
\end{tabular}
\label{tab:QCD-VR}
\end{table}

\begin{figure}[!htb]
\centering
\includegraphics[width=0.8\textwidth]{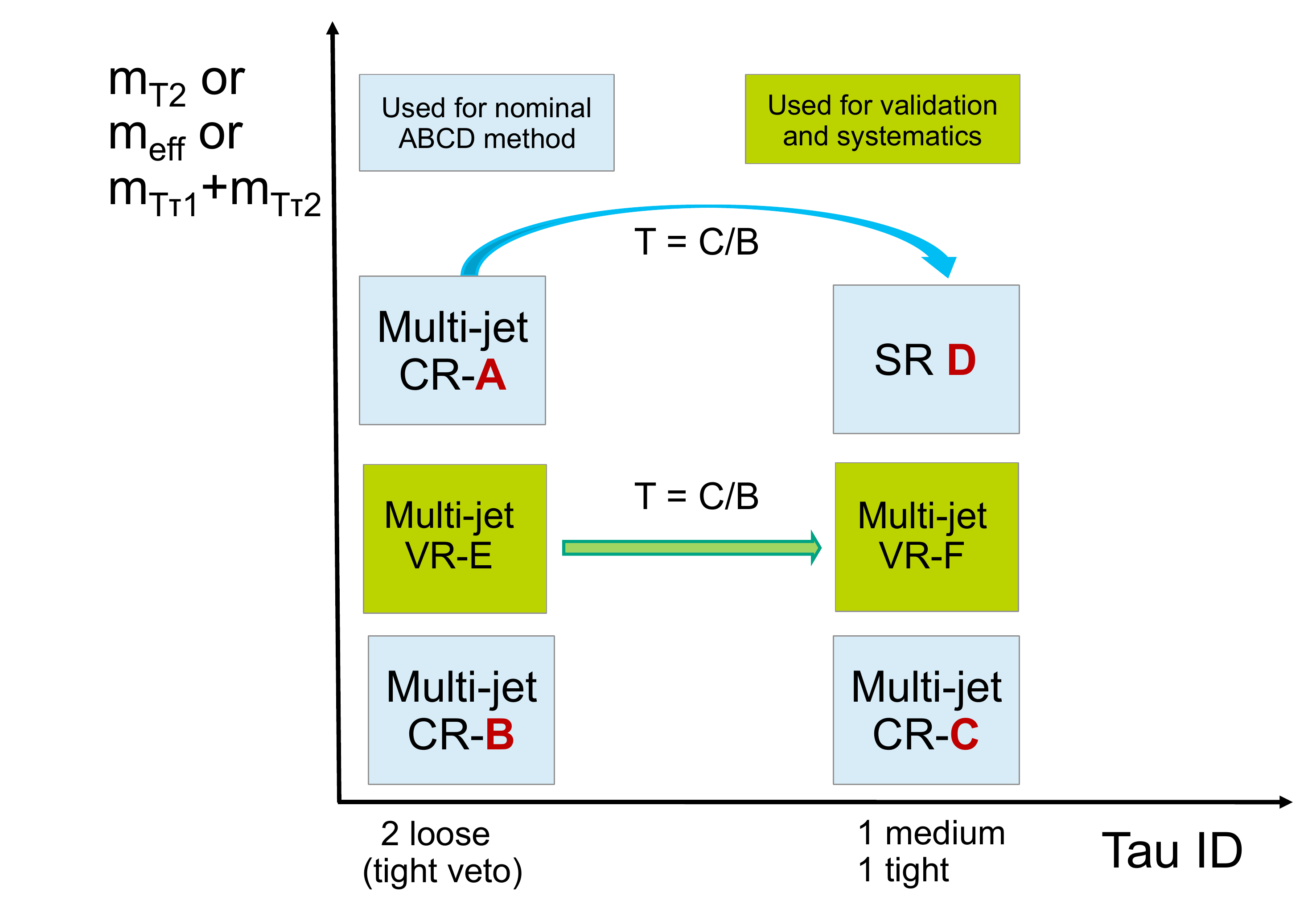}
\caption{ Illustration of the ABCD method for the multi-jet background determination.
The control regions A, B, C, and signal region D for the ABCD method described in the text 
(labelled as Multi-jet CR-A/B/C and SR D) are drawn as light blue boxes. 
Shown in green and labelled as Multi-jet-VR are the regions E and F, 
which are used to validate the ABCD method and to estimate the systematic uncertainties.}
\label{fig:abcd}
\end{figure}

The number of multi-jet events in the control and validation regions is estimated from data after
subtraction of other SM contributions estimated from MC simulation. 
Over 80\% of the events contributing to the control regions B and C come from multi-jet production.  
In control region A the multi-jet purity ranges from 60\% to 70\%, except for SR-DS-highMass for which the purity is 35\%.
The prediction in this region is affected by a large
statistical uncertainty.

The distributions of the kinematic variables in the control regions A, B and in validation region E are shown in figure \ref{fig:MET-regionAB}.
The results of the ABCD method are summarised in table
\ref{tab:qcd-est-result}. The SM predictions are in agreement with the observed data counts in the multi-jet
validation regions, as shown in table \ref{tab:qcd-valid-result}.

\begin{figure}[htpb]
\centering
\subfigure[]{    \includegraphics[width=0.45\textwidth]{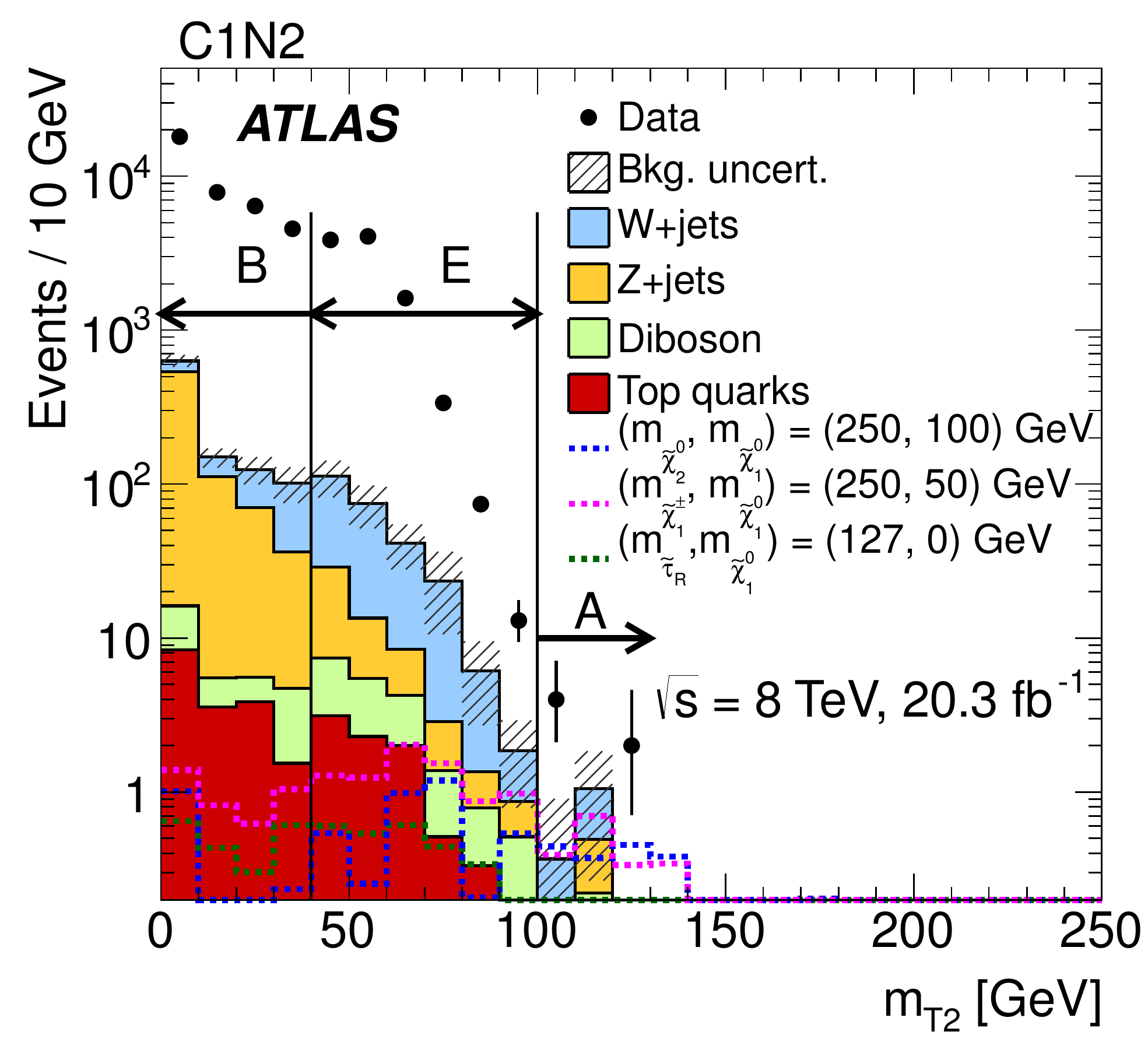}  }
\subfigure[]{    \includegraphics[width=0.45\textwidth]{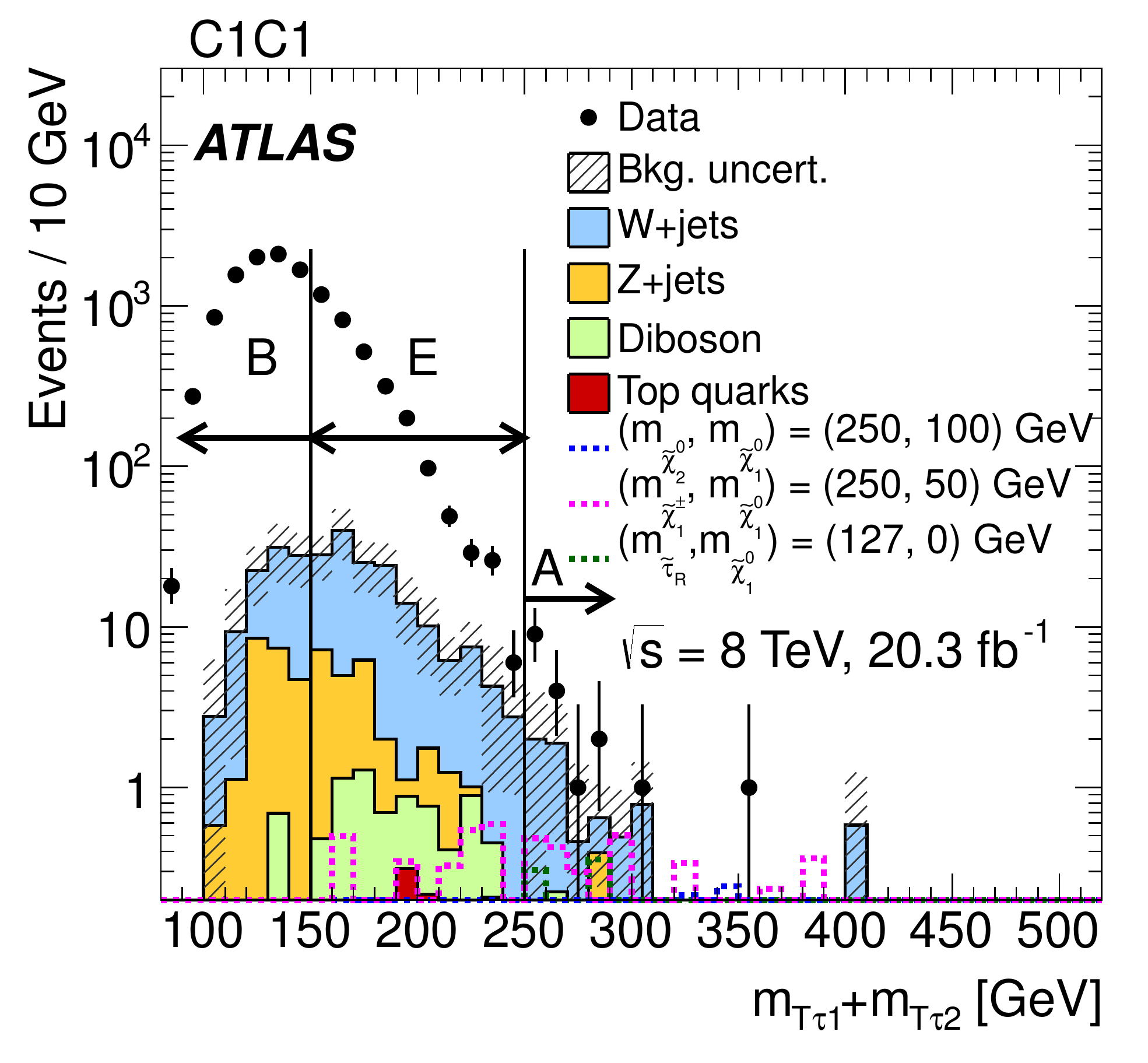}  }\\
\subfigure[]{   \includegraphics[width=0.45\textwidth]{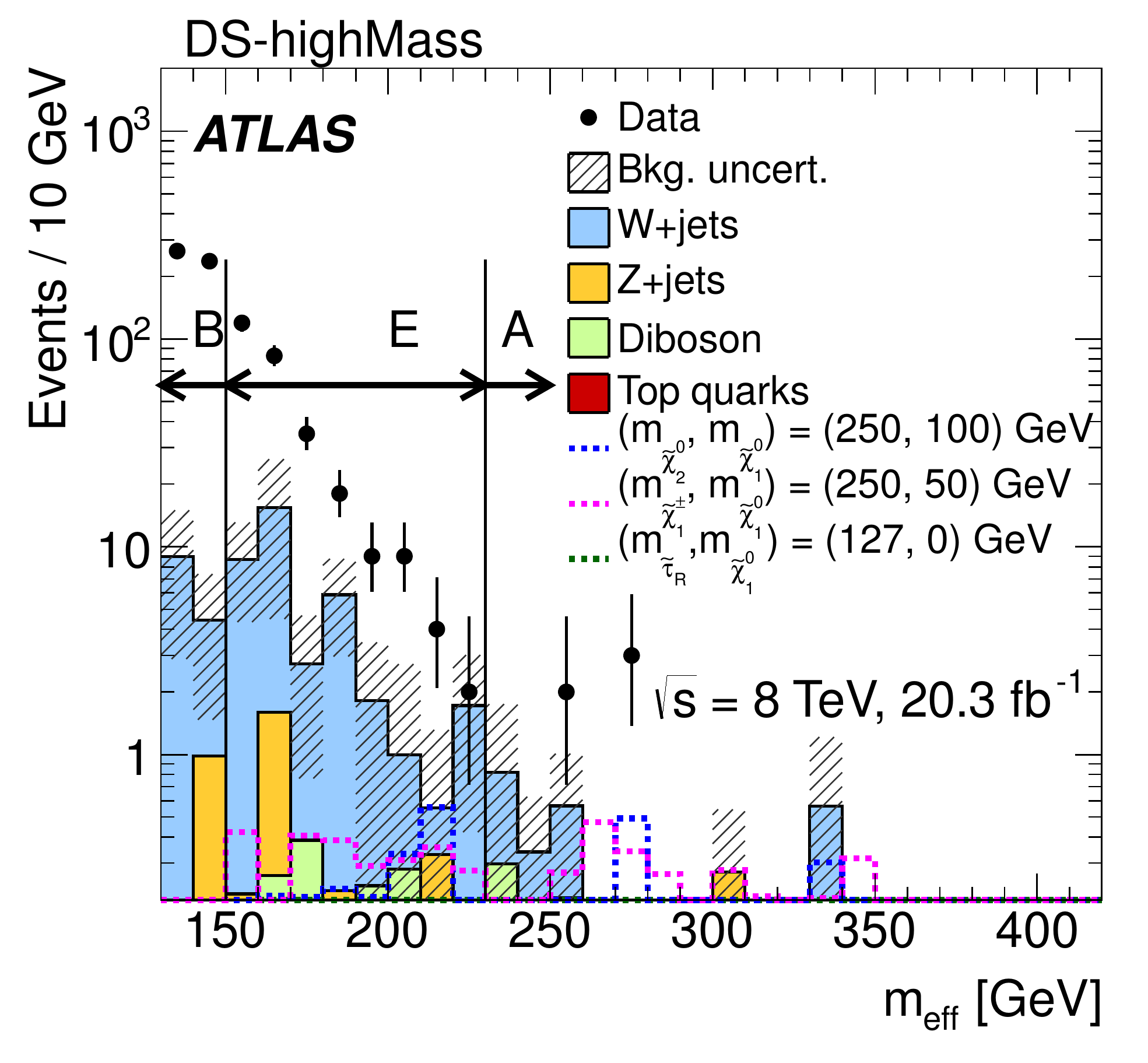}  }
\subfigure[]{  \includegraphics[width=0.45\textwidth]{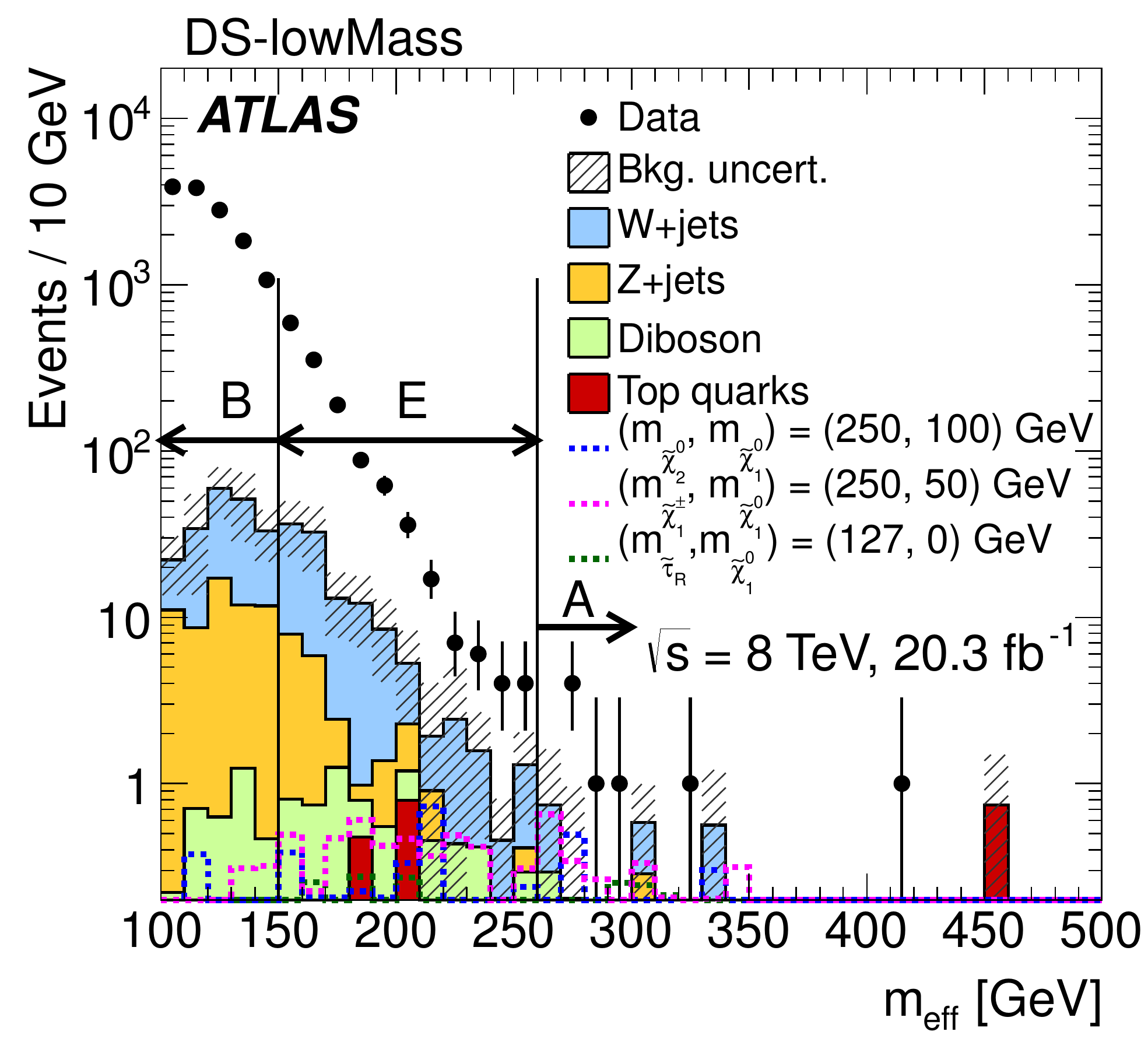}  }

\caption{(a) \mttwo, (b) $m_{{\rm T}\tau1}+m_{{\rm T}\tau2}$ and 
(c,d) $m_{\rm eff} $ distributions in the multi-jet background CRs A and B and in the
  validation region E defined in tables \ref{tab:QCD-CR} and \ref{tab:QCD-VR}.
The stacked histograms show the contribution of the non-multi-jet SM backgrounds from 
MC simulation, normalised to 20.3~\ifb. The hatched bands represent
the combined statistical and systematic uncertainties on the sum of the SM backgrounds shown. 
For illustration, the distributions of the SUSY reference points (see
section \ref{sec:signalGrids}) are also shown as dashed lines.
}
\label{fig:MET-regionAB}
\end{figure}

\begin{table}
\centering
\caption{
The MC predicted backgrounds in the multi-jet control regions, including both 
the statistical and systematic uncertainties,  and the expected multi-jet contribution (in italics), 
obtained by subtracting the MC contributions from observed data (in bold).  
Predicted event yields for the SUSY reference points in the control regions are also shown.
The estimated multi-jet contribution in the SRs is given in the last column.
The details of the systematic uncertainties reported here are
discussed in section \ref{sec:Systematics}. 
}
\footnotesize
\smallskip
\begin{tabular}{c|c|c|c|c|c|c}\hline
             &  Sample      & Region A          & Region  B       & Region  C     & $\rm{T}$ = C/B	  & Multi-jet in SR (D)\\ \hline \hline
             &  $\bold {Data }$        &  $\bold {6} $              &    $\bold {36907}$       & $\bold {24601}$           &		  &   \\ \cline{2-5}
             &  $Z$+jets    &  0.28 $\pm$ 0.16  & 730 $\pm$ 260   & 3980 $\pm$ 1060  &		  &   \\ \cline{2-5}
             &  $W$+jets    &  1.0 $\pm$ 0.4   & 250 $\pm$ 82   & 590 $\pm$ 180    &		  &   \\ \cline{2-5}
C1N2          &  diboson     & 0.51 $\pm$ 0.26    & 14.6 $\pm$ 4.8 & 72 $\pm $ 20    & 0.55	  & 2.3    \\\cline{2-5}
             & top          &   0.10 $\pm$ 0.06 & 17.3 $\pm$ 6.1 & 68 $\pm$ 22  &  $\pm$ 0.03  &  $\pm$ 1.4 \\ \cline{2-5}
             & \it {multi-jet }    & \it {4.1 $\pm$ 2.5 }     & \it {35900 $\pm$ 330 }& \it {19890 $\pm$ 1090 } & \\ \cline{2-5}
             & Ref. point 1  &  1.9 $\pm$ 0.9   &  1.4 $\pm$ 0.7 & 17.8 $\pm$ 6.2  &		  &  \\ \hline \hline  
             & $\bold {Data }$         &  $\bold {18}$          &   $\bold {8479}$     &  $\bold {4551}$     &		  &   \\ \cline{2-5}
             &  $Z$+jets    &  0.06 $\pm$ 0.06 &   21 $\pm$ 10   & 80 $\pm$ 25    &		  &   \\ \cline{2-5}
             &  $W$+jets    &  5.6 $\pm$ 1.2   & 71 $\pm$ 32     & 160 $\pm$ 46    &		  &   \\ \cline{2-5}
 C1C1   &  diboson     & 0.9 $\pm$ 0.4    & 1.2 $\pm$ 0.6  & 5.4 $\pm$ 2.2   & 0.51	  & 5.8   \\ \cline{2-5}
             & top          & 0.11 $\pm$ 0.06  &   0.0 $\pm$ 1.0       & 0.6 $\pm$ 0.5   &  $\pm$ 0.01  &   $\pm$ 2.3 \\ \cline{2-5}
             & \it {multi-jet }    & \it {11.3 $\pm$ 4.4 }  & \it {8390 $\pm$ 98 }  &  \it {4300 $\pm$ 85 }  &		  & \\ \cline{2-5}
             & Ref. point 2 &  3.9 $\pm$ 1.0   &  0.13 $\pm$ 0.09 &  1.4 $\pm$ 0.4  &		  &  \\ \hline\hline
             & $\bold {Data }$         &  $\bold { 5}$         &   $\bold {500}$      &    $\bold {268 }$         &		  & \\ \cline{2-5}
             & $Z$+jets     & 0.24 $\pm$ 0.18   & 1.0 $\pm$ 0.7  &  1.7 $\pm$ 1.0  &		  & \\ \cline{2-5}  
             &  $W$+jets    &  2.2 $\pm$ 0.7   & 12.3 $\pm$ 6.6    & 20.6 $\pm$ 9.6      &		  &   \\ \cline{2-5}
 DS-      & diboson      &    0.7 $\pm$ 0.4 &  0.2 $\pm$ 0.1  & 0.6 $\pm$ 0.3   & 0.50	  & 0.9 \\ \cline{2-5}  
highMass      & top          &    0.06 $\pm$ 0.03 & 0.0 $\pm$ 1.0  &  0.8 $\pm$ 0.6  &  $\pm$ 0.05  & $\pm$1.2 \\ \cline{2-5}  
             & \it {multi-jet }   &    \it {1.8 $\pm$ 2.4 } &  \it {487 $\pm$ 23 }  & \it { 244 $\pm$ 19 }  &		  &  \\ \cline{2-5}
             & Ref. point 3 &    0.9 $\pm$ 0.4 &  0.0 $\pm$ 1.0 &  0.20 $\pm$ 0.13  &		  & \\  \hline \hline 
             & $\bold {Data }$            &         $\bold {8 }$                &     $\bold {13419}$     &     $\bold { 7632}$       &		  & \\ \cline{2-5}
             & $Z$+jets     &  0.14 $\pm$ 0.09  &   57 $\pm$ 26  &  180 $\pm$ 49   &		  &  \\ \cline{2-5}
             & $W$+jets     &  2.3 $\pm$ 0.7   & 140 $\pm$ 51   & 290 $\pm$ 75    &		  &   \\ \cline{2-5}
DS-  & diboson     &   0.40 $\pm$ 0.24 &  3.1 $\pm$ 1.4 &  10.7 $\pm$ 3.6 &  0.54	  &  2.8 \\ \cline{2-5}
lowMass  & top          & 0.09 $\pm$ 0.05   &  0.1 $\pm$ 0.1 &  3.8 $\pm$ 1.9  &   $\pm$ 0.01 & $\pm$ 1.7\\ \cline{2-5} 
             & \it {multi-jet }    &  \it {5.1 $\pm$ 2.9 }  &\it {13220 $\pm$ 130 } &  \it {7150 $\pm$ 130 } &		  &  \\ \cline{2-5}
             & Ref. point 3 &  1.4 $\pm$ 0.5  &  0.11 $\pm$ 0.11 &   1.3 $\pm$ 0.4 &		  & \\  \hline

\end{tabular}

\label{tab:qcd-est-result}
\end{table}

\begin{table}
\caption{Number of events in the multi-jet validation regions F for data and SM backgrounds, including both statistical and systematic
  uncertainties.  The SM MC backgrounds are normalised to  20.3~\ifb. The multi-jet contribution is estimated
  from data with the ABCD method, by multiplying  the number of events in region E and the transfer
factor $\rm{T}$. 
}
\centering
\smallskip
\begin{tabular}{c|c|c|c|c}\hline
 Sample   & VR-C1N2        & VR-C1C1            & VR-DS-highMass   & VR-DS-lowMass          \\ \hline    
 Data               &  5585              &  1846                   & 163                       &  764 \\ 
 SM total         & $5840 \pm 340$ & $1870 \pm 87$ 	& $170 \pm 24$& $860 \pm 110$\\ \hline 
 Multi-jet        &$5370 \pm 320$ &$1570 \pm 61 $  	&$120 \pm 19 $ &$670 \pm 100 $ \\ 
  $W$+jets       & 320 $\pm$ 93  & 240 $\pm$ 60  & 37 $\pm$ 15 & 130 $\pm$ 42 \\ 
 $Z$+jets       & 97 $\pm$ 39     & 34 $\pm$ 12  & 5.6 $\pm$ 2.9 & 36 $\pm$ 14 \\ 
 Diboson         & 27 $\pm$ 8  & 15.1 $\pm$ 5.4  & 5.1  $\pm$ 1.5 & 12.9 $\pm$ 3.6 \\ 
 Top                 & 24 $\pm$ 10 & 5.0 $\pm$ 2.1    &  3.6 $\pm$ 1.7 & 9.5 $\pm$ 3.3 \\ \hline
\end{tabular}
\label{tab:qcd-valid-result}
\end{table}

\subsection{$W$+jets background estimation}
\label{sec:BG_W}

The production of $W$+jets events with at least one misidentified tau is an important background in the SRs, ranging from 25\% to 50\%.
A dedicated control region ($W$~CR) is used to normalise the $W$+jets MC estimation to data.
The $W$~CR is designed to be kinematically as close as possible to the signal regions, and is enriched in events where the $W$ decays 
leptonically into a muon and a neutrino to suppress multi-jet contamination.
Events containing exactly one isolated muon and one tau passing the tight jet BDT requirement are selected.
The selected leptons must have opposite electrical charge.
To reduce the contribution from $Z$+jets production, $m_{{\rm T},\tau} + m_{{\rm T},\mu} >$~80~GeV is required, and the reconstructed invariant 
mass of the two leptons, $m_{\tau, \mu}$, must be outside the $Z$ mass window (12~GeV~$<m_{\tau, \mu}<$~40~GeV or $m_{\tau, \mu}>$~100~GeV).
To further suppress multi-jet and $Z$+jets events, \met~$>$~40~GeV is required, and the leptons must not be back-to-back, i.e. 
$|\Delta\phi(\tau, \mu)|<$~2.7 and $|\Delta\eta(\tau, \mu)| <$~2.0. 
The contribution from events with top quarks is suppressed by rejecting events containing $b$-tagged jets, i.e. $N$(\verb|B20|) = 0.
The definition of the $W$~CR is given in  table \ref{tab:1tau_WCR}.

The multi-jet contribution in the $W$~CR is estimated by counting the number of events in data 
satisfying the same requirements as the $W$~CR but with same-sign (SS) charge of the two leptons.
Events from SM processes other than multi-jet production are subtracted from
the data counts in the SS region, using their MC prediction.
The method relies on the fact that in the multi-jet background the ratio of SS to OS events is close to unity, whilst a significant difference from 
unity is expected for $W$+jets production. The latter is dominated by $gu/gd$-initiated processes that often give rise to a jet originating from a quark, the 
charge of which is anti-correlated with the $W$ boson charge.

The $W$+jets estimate is tested in a validation region ($W$~VR) where the $W$ is required to decay leptonically into an electron and a neutrino.
The $W$~VR is defined in the same way as the $W$~CR except that events with one tight tau and one isolated electron are selected. 
The multi-jet contribution in the $W$~VR is estimated using the same technique as in the $W$~CR. 
The event yields in the $W$+jets control and validation regions are given in table \ref{tab:WCR}. 
The purity of the selection in $W$+jets events is around 80\% (75\%) in the $W$~CR (VR).  Good agreement between data and SM predictions is observed. 
The large uncertainties stem from the corrections applied for the trigger and identification efficiency for misidentified taus from light jets.
Distributions of the kinematic variables defining the SRs are shown in figure \ref{fig:WCR}. The distribution of the effective mass $m_{\rm eff}$ in the $W$ 
VR is shown in figure \ref{fig:bkg-vr} (a).

\begin{table}[h]
  \caption{Definition of the $W$+jets control region.}
\smallskip
  \centering
  \begin{tabular}{c}
    \hline
    1 tight tau and 1 isolated muon with opposite charge    \\
    $b$-jet veto                             \\
    $\Delta\phi(\tau, \mu) < 2.7$             \\
    $|\Delta\eta(\tau, \mu)| < 2.0$           \\
    \met~$>40$ GeV                       \\
    $m_{{\rm T},\tau} + m_{{\rm T},\mu} > 80 \GeV$         \\
12~GeV~$<m_{\tau, \mu}<$~40~GeV or $m_{\tau, \mu}>$~100~GeV\\
    \hline
    \end{tabular}
  \label{tab:1tau_WCR}
\end{table}

\newcommand{\Wjetsx}{$W+$jets}
\newcommand{\Zjetsx}{$Z+$jets}
\begin{table}[!h]
  \caption{
    Event yields in the $W$+jets control and validation regions. 
    The SM backgrounds other than multi-jet production are estimated from MC simulation and normalised to 20.3~\ifb.
    The multi-jet contribution is estimated from data in the same-sign region, by subtracting all other SM backgrounds from the data counts.
    The quoted uncertainty is the sum in quadrature of statistical and systematic uncertainties.
  }
  \smallskip
  \centering
  \begin{tabular}{c|c|c}
  
    \hline
    Sample   & $W$~CR                     & $W$~VR \\

\hline
Data       & 4120 
           & 3420  
           \\
\hline
SM total   & $ 4100 \pm   900$  
           & $ 3500 \pm   600$  
\\
\hline
\Wjetsx    & $ 3300 \pm   800$  
           & $ 2600 \pm   500$  
\\
Top        & $  250 \pm    80$  
           & $  240 \pm    70$  
\\
Diboson    & $  180 \pm    50$  
           & $  170 \pm    40$  
\\
\Zjetsx    & $  140 \pm    40$  
           & $   99 \pm    31$  
\\
Multi-jet  & $  250 \pm   250$  
           & $  400 \pm   200$  
\\
\hline
\end{tabular}
\label{tab:WCR}
\end{table}

\begin{figure}
  \centering
  %
\subfigure[]{    \includegraphics[width=0.45\textwidth]{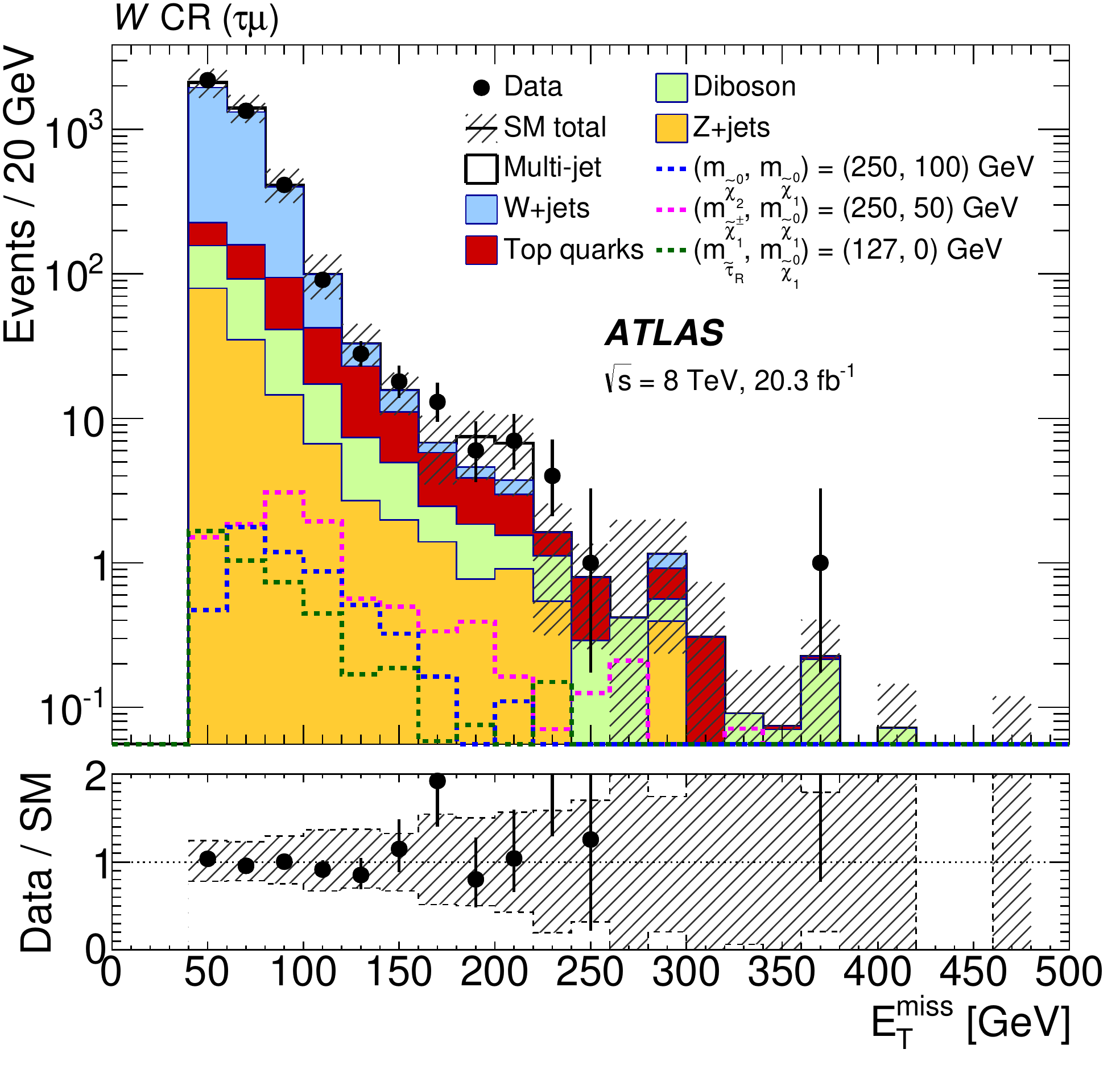}  }   
\subfigure[]{    \includegraphics[width=0.45\textwidth]{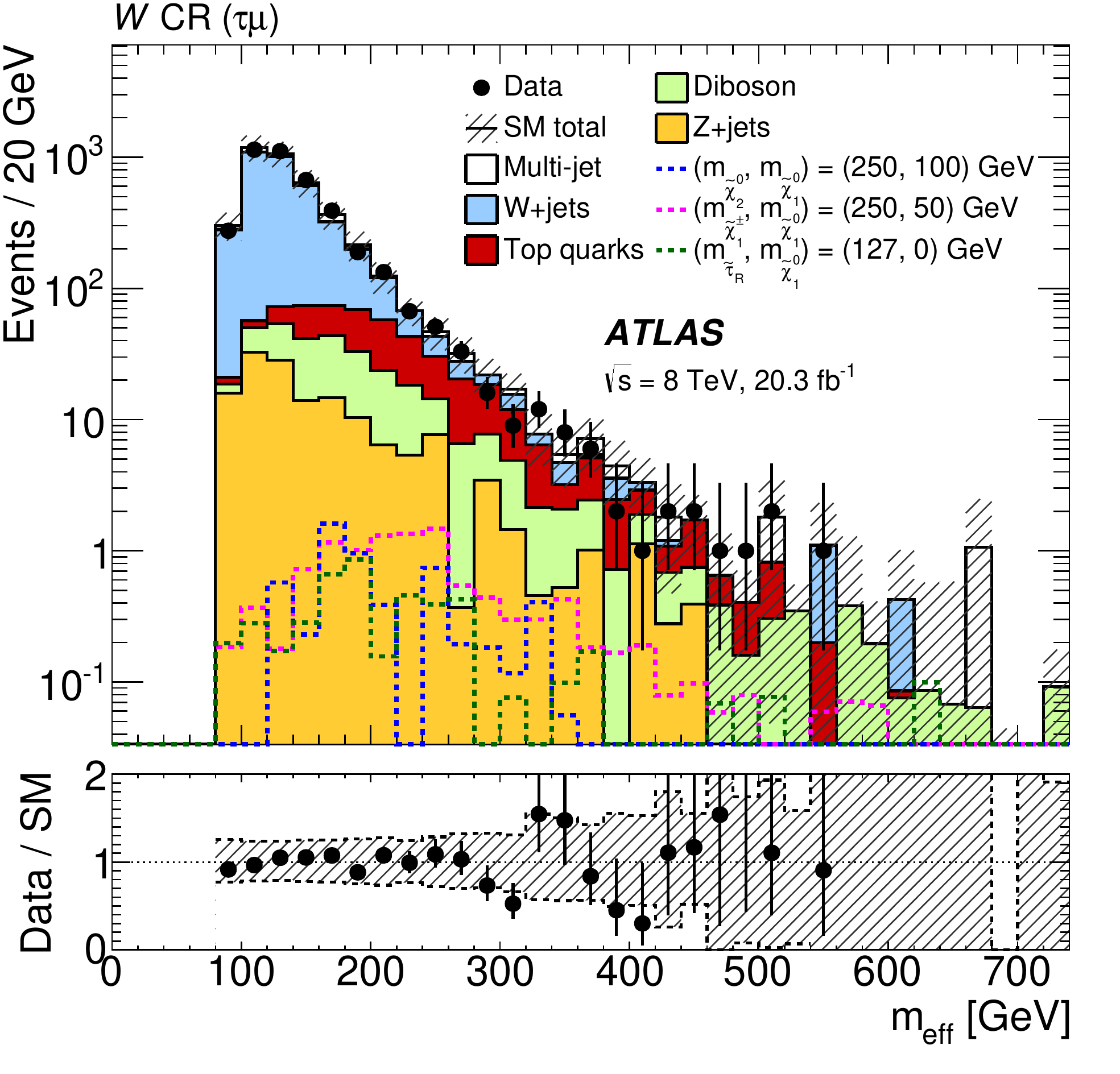}  }   
\\
\subfigure[]{    \includegraphics[width=0.45\textwidth]{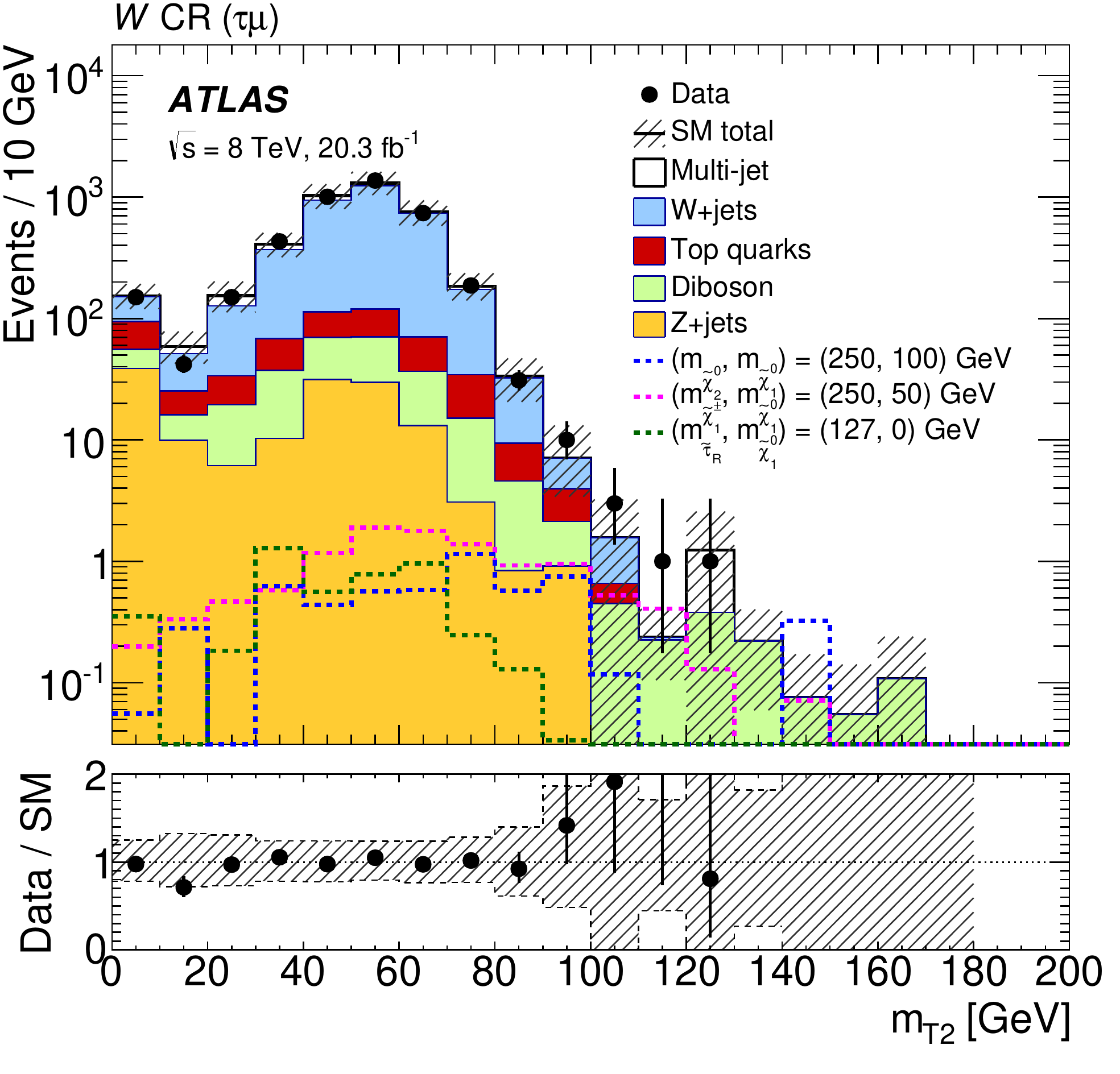}  }   
\subfigure[]{    \includegraphics[width=0.45\textwidth]{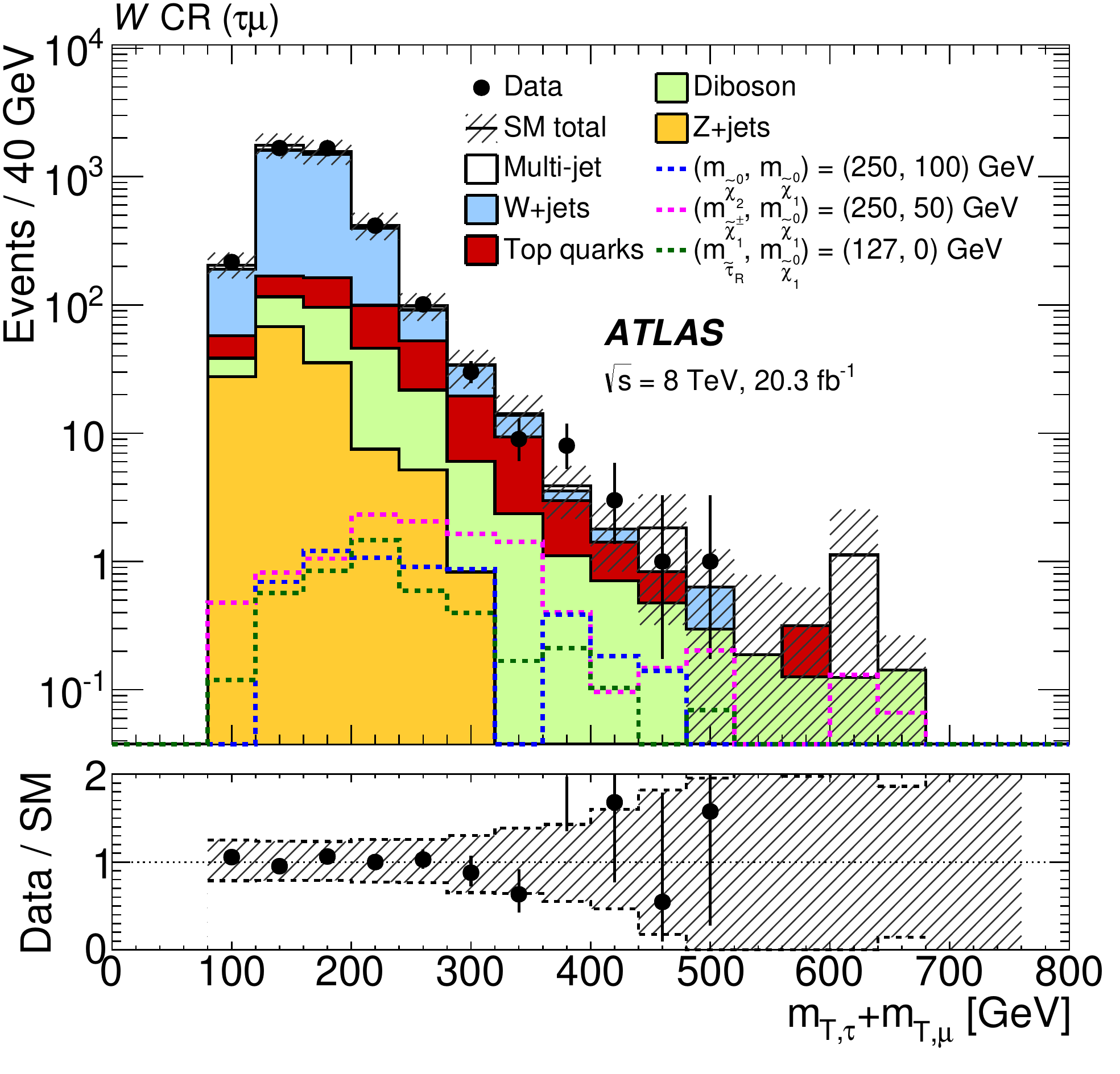}  }   
  %
  \caption{
    Distributions in the $W$+jets control region of the kinematic variables which are used in the signal region definition.    
    The SM backgrounds other than multi-jet production are estimated from MC simulation and normalised to 20.3~\ifb.
    The multi-jet contribution is estimated from data in the same-sign region, by subtracting all other SM backgrounds from the data counts.
    The hatched bands represent the combined statistical and
    systematic uncertainties on the total SM background. For
    illustration, the distributions of the SUSY reference points are
    also shown as dashed lines. The lower panels show the ratio of data to the SM background estimate. 
  }
  \label{fig:WCR}
\end{figure}

\subsection{Estimation of irreducible backgrounds}
\label{sec:BG_irre}

Irreducible SM backgrounds arise mainly from \ensuremath{t\bar{t}}, single top,
\ensuremath{t\bar{t}}+$V$, $Z/\gamma^{*}$+jets and diboson ($WW$, $WZ$ and $ZZ$) processes and
are estimated with MC simulation and validated in data. Other SM backgrounds are negligible.

The diboson background accounts for 15--35\% of the total SM
contribution in the signal regions, and mainly arises 
from $WW\rightarrow \tau\nu \tau\nu$  and $ZZ \rightarrow \tau\tau \nu\nu$ events. 
To validate the MC modelling and normalisation of the $WW$ process, validation regions with enriched $WW
\rightarrow e \nu \mu\nu$ contribution are defined for each SR. Events with exactly one OS electron--muon pair are selected. 
To keep the same phase space as the SRs, the $WW$  VRs are defined in the same
way except for intermediate requirements on the relevant kinematic variable to
reduce possible signal contamination, 
i.e. 50~GeV~$<\mttwo<$~100~GeV for SR-C1N2, 150~GeV~$< m_{{\rm T}, \mu}+m_{{\rm T}, e}<$~250~GeV for SR-C1C1, 
150~GeV~$<m_{\rm eff}<$~230~GeV for SR-DS-highMass and 150~GeV~$<m_{\rm eff}<$~260~GeV for SR-DS-lowMass.
The purity of the selection in $WW$ events ranges from 70\% to 80\% in all validation regions. 
The agreement between data and SM predictions is found to be good, and the data
yields match the SM predictions within uncertainties (of the order of
10\%), as shown in figure \ref{fig:bkg-vr} (c).

The inclusive contribution from \ensuremath{t\bar{t}}, single top, \ensuremath{t\bar{t}}+$V$ and $Z/\gamma^{*}$+jets
amounts to about 5--20\% of the total background in the signal regions.
The MC estimates are validated in regions enriched in $Z$+jets and \ttbar~ events. 

In the $Z$+jets validation region events containing one isolated electron or muon and one tau with opposite electrical charge are selected.
The invariant mass of the dilepton pair must satisfy 40~GeV~$< m_{\ell\tau} <$~75~GeV. 
To suppress contributions from other SM processes, $m_{{\rm T},
  l}+m_{{\rm T}, \tau}
<$~80~GeV, $|\Delta\phi(\ell,\tau)| >$ 2.4 and $N$(\verb|B20|) = 0 veto are
required. The multi-jet contribution in the $Z$ VR is estimated with the same
method as in the $W$~CR and VR.
 
The top quark validation region is defined by requiring at least two candidate taus, and at least one must satisfy the tight jet BDT quality 
requirement. At least one of the selected tau pairs must contain taus with opposite electrical charge.
To increase the contribution from top events, two $b$-tagged jets with
$p_{\rm T} > 20$ GeV are required, and the events should be
kinematically compatible with \ttbar~production (top-tagged) 
through the use of the variable $m_{\rm {CT}}$ \cite{Tovey:2008}.
The scalar sum of the $p_{\rm T}$ of the two taus and of at least one
combination of two jets in an event must exceed 100 GeV. Furthermore, top-tagged events are required to possess $m_{\rm {CT}}$
values calculated from combinations of jets and taus consistent with the expected bounds from $t\bar{t}$ events as described in
Ref. \cite{Polesello:2009rn}. 
Requirements of 50~GeV~$<$~\met~$<$~100 GeV and $m_{{\rm T}\tau1}+m_{{\rm T}\tau2} >$~80~GeV are
applied to suppress contributions from SM backgrounds not containing top quarks. 

The purity of the selection in $Z$+jets and \ttbar~ events is above 80\% in the respective
validation regions and good agreement between the data and SM expectation is observed.
Distributions of relevant kinematic variables in the $Z$ and top VRs
are shown in figure \ref{fig:bkg-vr} (b) and (d).

\begin{figure}[htpb]
\centering
\subfigure[~$W$ VR] {  \includegraphics[width=0.45\textwidth]{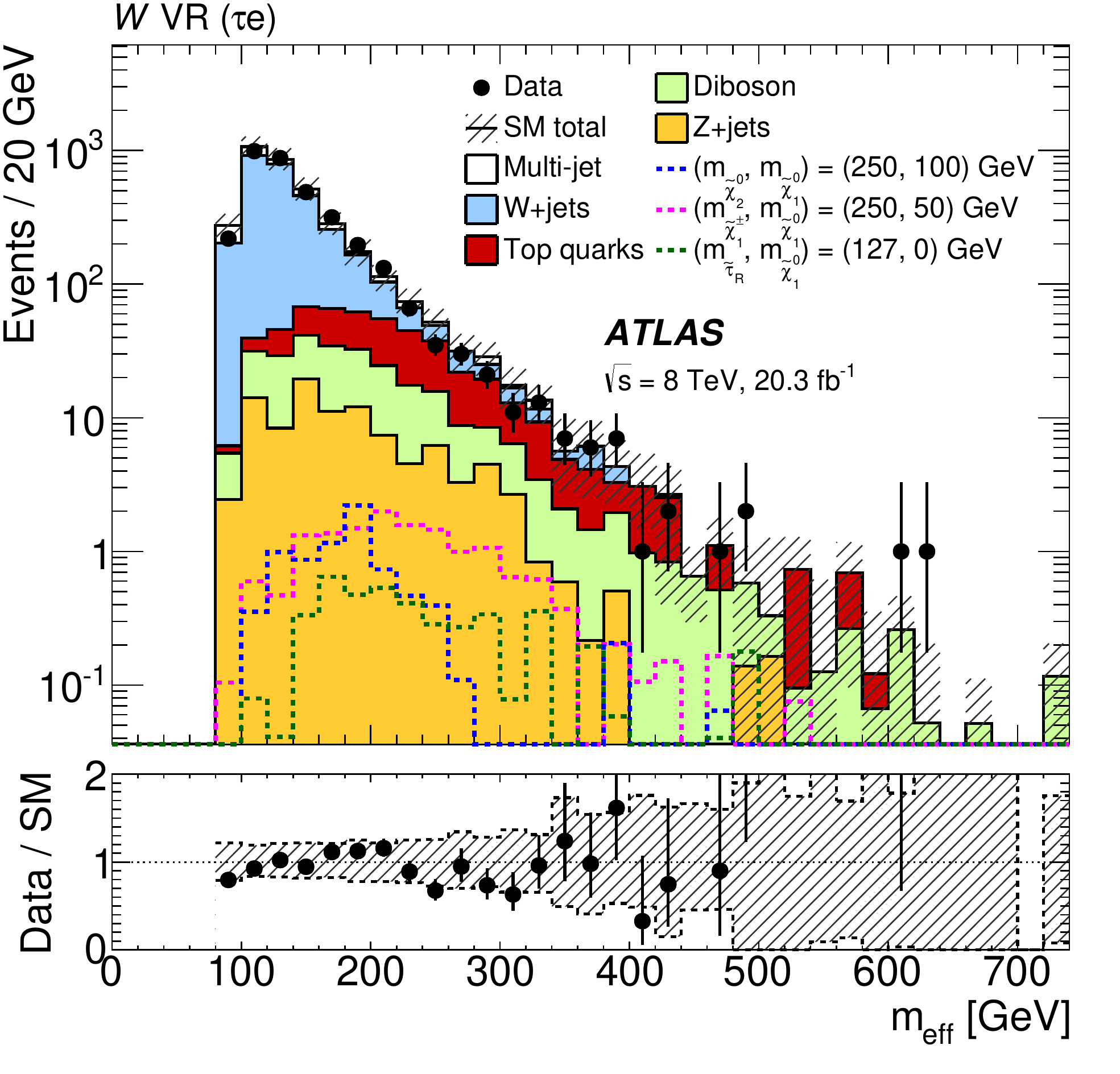}  }    
\subfigure[~$Z$ VR] {  \includegraphics[width=0.45\textwidth]{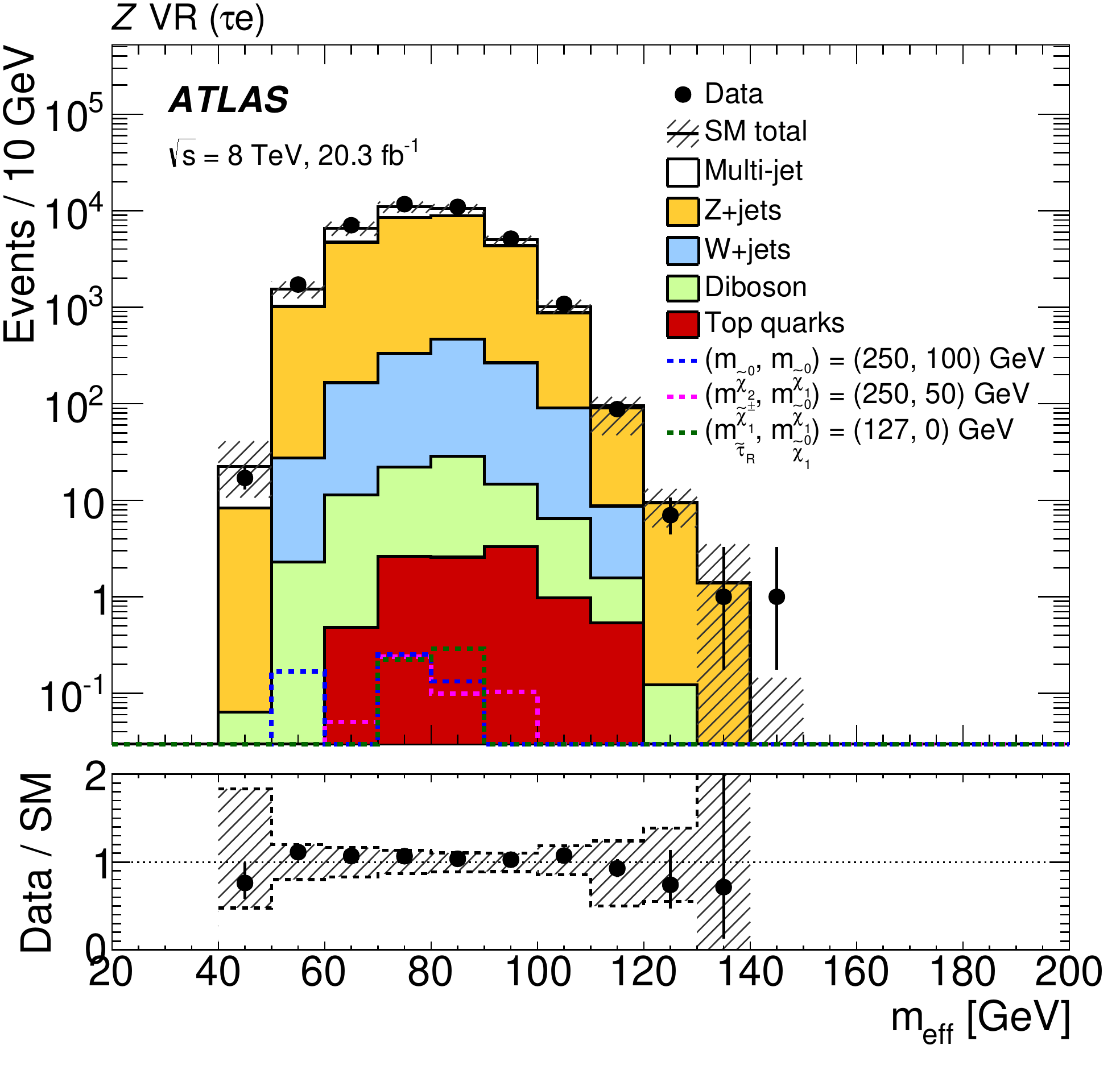}  }\\  
\subfigure[~$WW$ VR C1N2]{  \includegraphics[width=0.45\textwidth]{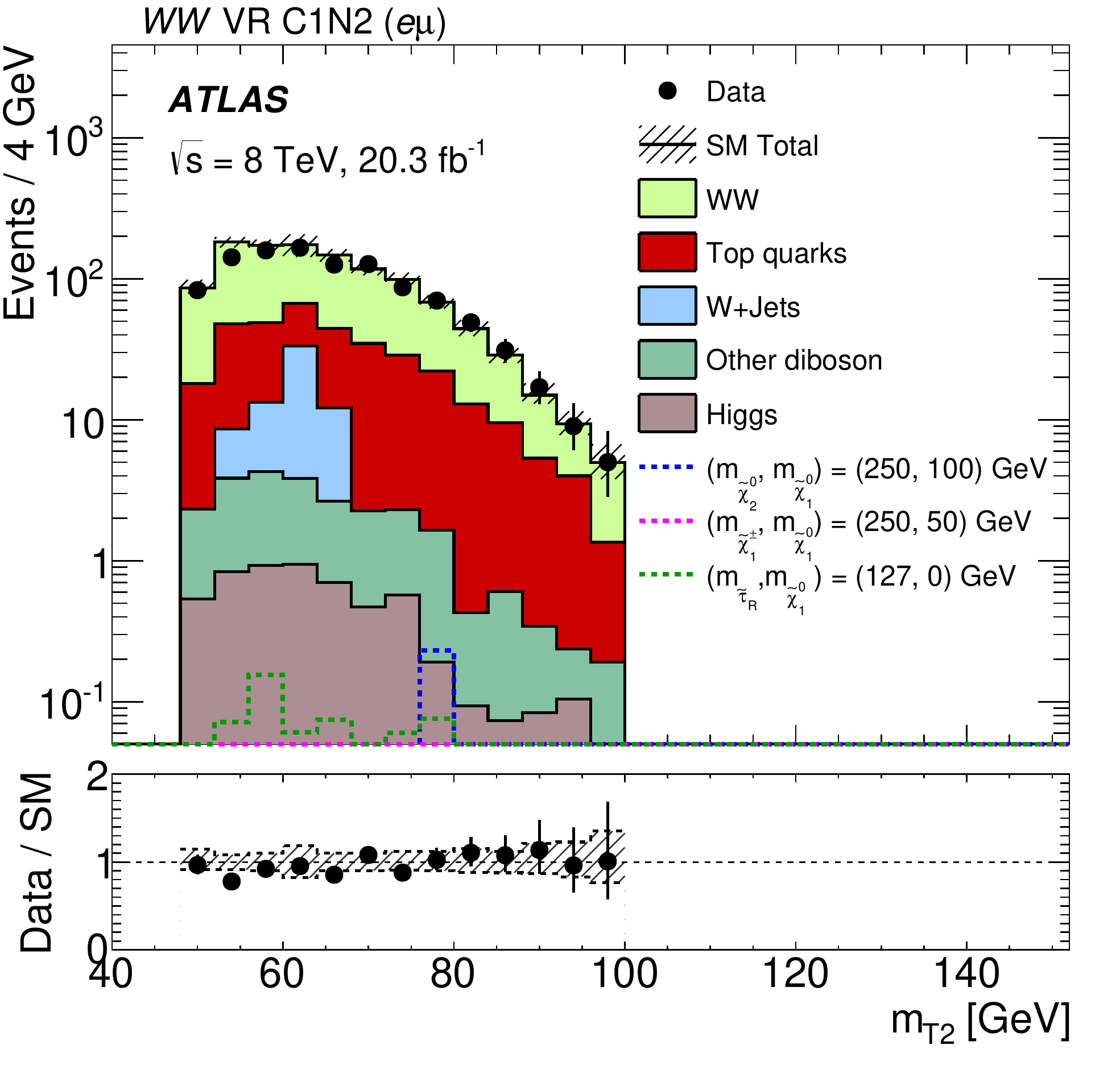}  }
\subfigure[~top VR] {  \includegraphics[width=0.45\textwidth]{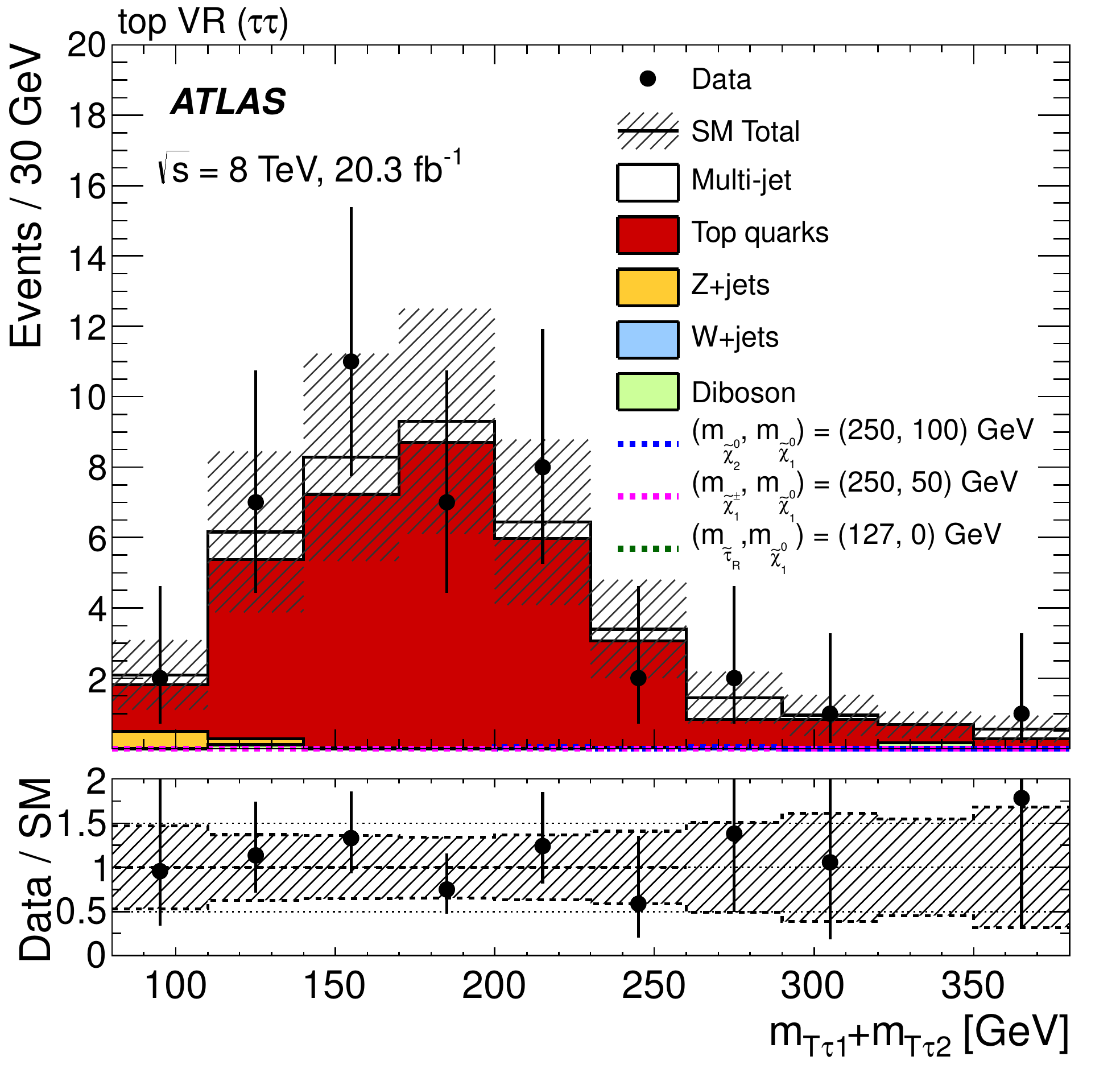}  }

\caption{Distributions of relevant kinematic variables in the
  validation regions: $m_\text{eff}$ in the (a) $W$ and (b) $Z$ VRs,
  (c) \mttwo\ in the $WW$ VR C1N2, 
and (d) $m_{{\rm T}\tau1}+m_{{\rm T}\tau2}$ in the top VR. In the $Z$ VR, events containing one electron and one tau are shown.
The SM backgrounds other than multi-jet production are estimated from MC simulation 
and normalised to 20.3~\ifb. The hatched bands represent the combined statistical and systematic uncertainties on the total SM background. 
For illustration, the distributions of the SUSY reference points are
also shown as dashed lines. The lower panels show the ratio of data to the SM background estimate. 
}
\label{fig:bkg-vr}
\end{figure}

\subsection{Fitting procedure}
\label{sec:BG_fit}

For each signal region, a simultaneous fit is performed based on the profile likelihood method \cite{statforumlimitsl}. The inputs to the
fit are: 
the number of observed events in the multi-jet CR A and $W$~CR,
the expected non-multi-jet and non-$W$ contributions to the multi-jet CR~A and $W$~CR,
the expected contributions of multi-jet and $W$ to the multi-jet CR~A
and $W$~CR,  as described in sections \ref{sec:BG_QCD} --
\ref{sec:BG_W}, 
and the transfer factors, which relate the number of multi-jet or $W$+jets events in their associated control region to that predicted in the signal region.
The number of events in a given CR is described using a Poisson distribution, 
the mean of which is the sum of the expected contributions from all background
sources. The free parameters in the fit are the normalisations of the $W$+jets and multi-jet contributions. The systematic uncertainties on
the expected background yields are included as nuisance parameters,
assumed to be Gaussian distributed with a width determined from the size of
the uncertainty. Correlations between control and signal regions, and
background processes are taken into account with common nuisance
parameters. The fit parameters are determined by maximising the
product of the Poisson probability functions and the constraints for
the nuisance parameters.

\section{Systematic uncertainties}
\label{sec:Systematics}

Systematic uncertainties have an impact on the estimates of the background and signal event yields
in the control and signal regions.  
Several sources of systematic uncertainty are considered for the ABCD method used to 
determine the multi-jet background: the correlation between the kinematic variable (\mttwo, 
$m_{{\rm T}\tau1}+m_{{\rm T}\tau2}$, or $m_{\rm eff}$) and the tau identification variable, the
limited number of events in the CRs, and the subtraction of other SM 
backgrounds.
The systematic uncertainty on the correlation is estimated by comparing the transfer
factor from the region  B to region C to that of the region E to F (see figure \ref{fig:abcd}).
The systematic uncertainty due to the limited number of events in the control regions is estimated by 
considering the statistical uncertainty on the number of data events
and the other SM background components. 
The systematic uncertainty on the non-multi-jet background subtraction in the
control regions A, B, and C is estimated by considering the systematic
uncertainty in the MC estimations of the non-multi-jet background in the CRs.
A summary of the systematic uncertainties on the multi-jet background estimation in the SRs is shown in table \ref{tab:syst_QCD}.
The dominant source is given by the limited number of events in the multi-jet control regions.

The dominating uncertainty in the multi-jet estimation from SS events in the $W$ control region and in the $W$ and $Z$ validation regions stems from the
different fractions of quark and gluon jets in the OS and SS regions. The relative difference in the tau mis-identification rates measured in events with OS and 
SS lepton pairs varies between 20\% and 70\% in most of the phase space of interest for this analysis; however, it goes up to 100\% in regions with few events.

\begin{table}[h]
\begin{center}
\footnotesize
\caption{\label{tab:syst_QCD} Summary of the systematic uncertainties
  for the multi-jet background estimation. The total uncertainty is
  the sum in quadrature of each source.} \smallskip
\begin{tabular}{c|c|c|c|c}
\hline
Systematic  Source                                       & SR-C1N2           &SR-C1C1      & SR-DS-highMass & SR-DS-lowMass  \\
\hline
Correlation                                           &    4.9\%       &  1.6\%      &   8.0\%      &  14\%     \\
Non-multi-jet subtraction in Region A	    &	   8.0\%     &12\%     &    21\%     &   13\%	 \\
Non-multi-jet subtraction in Region B	    &	   1.0\%	& 0.4\%	&  1.2\%	&   0.5\%	\\
Non-multi-jet subtraction in Region C	    &	  2.7\%	&  1.4\%  &  3.6\%       &   2.0\%      \\
Number of events in Region A          &    61\%         &38\%     &    133\%       &   57\%   \\
Number of events in Regions C and B   &     1.0\%        & 2.0\%  &     8.4\%       &   1.5\%   \\
\hline
Total                                                     &     62\%     &   40\% &    135\%         &  60\%  \\
\hline     
\end{tabular}
\end{center}
\end{table}

The experimental systematic uncertainties on the SM backgrounds estimated with MC simulation are due to
those on the jet energy scale and resolution \cite{Aad:2011he}, \met\ energy scale and resolution \cite{Aad:2012re},
$b$-tagging and mis-identification efficiency~\cite{bperf1},
and tau identification, trigger efficiency and energy scale \cite{atlas-tau-energyscale}.
The main contribution comes from the uncertainty on the corrections applied to the tau identification
and trigger efficiency in the simulated samples, and from the tau energy scale uncertainty. 
The uncertainty on the correction for the tau identification efficiency is of the order of 2--3\% for taus with $p_{\rm T}$ between 20 and 200 GeV.
Uncertainties on the trigger efficiency are around 2--4\% for taus with $p_{\rm T}$ between 30 and 50 GeV, and increase to 8--10\% for taus with $p_{\rm T}>$~50~GeV. The uncertainty on the tau energy scale is of the order of a few percent, with no significant dependence on pile-up or on the tau pseudorapidity.  

A systematic uncertainty  of the order of 10\% associated with the simulation of pile-up is also taken into account.
The luminosity uncertainty is 2.8\%~\cite{lumi2012}.

Theoretical uncertainties affecting the generator predictions arise
from the finite number of partons (FNP) from the hard 
primary interaction for diboson and top processes, the effect of the renormalisation ($\mu_{{\rm R}}$) and
factorisation ($\mu_{\rm F}$) scales for diboson, $W$+jets and $Z$+jets production, and the impact of the merging scale of the matrix 
element with the parton shower (ME-PS) for diboson and top processes. The FNP-induced uncertainty is calculated by comparing the baseline
samples which contain up to three additional partons from the hard interaction 
to samples with up to two additional partons. The uncertainty due to the $\mu_{{\rm R}}$ and $\mu_{{\rm F}}$ scales is determined by the comparison with 
 samples with these scales varied up and down by a factor of two, while the ME-PS merging scale uncertainty is determined by a comparison with 
samples with varied merging scale. For $W$+jets events, the
uncertainty due to the jet \pt\ threshold used for parton--jet matching 
is calculated by comparing the baseline samples with jet \pt\ threshold set to 15 GeV to samples with a threshold of 25 GeV. 
The theoretical uncertainties associated with the top sample vary from 10\% to 20\% while the $W/Z+$jets
and diboson theoretical uncertainties vary from 17\% to 30\%,
depending on the process and the signal region.
The uncertainty due to the PDF choice is 1--2\% for $W/Z$+jets and diboson
production, and around 6\% for \ttbar, single top and \ttbar+$V$ processes.
The theoretical uncertainties on the cross sections are 5\% for $Z+$jets~\cite{ZcrossSection}, 
10\% for $t\bar{t}$, 3\%, 4\% and 7\% for single top in the $t$-, $s$- and $Wt$- channels, respectively,
 22\% for \ttbar+$W/Z$, and 5\%, 5\% and 7\% for $WW$, $ZZ$ and $WZ$, respectively~\cite{Campbell:2011bn}.

The various sources of uncertainty for the non-multi-jet background estimates in the SRs are summarised in table \ref{tab:syst_MC}.
In the $W$~CR, the total uncertainty on the $W$+jets estimation is around 20\%, and it is dominated by theory uncertainties and corrections due to
trigger and tau identification efficiency.

\begin{table}[h]
\begin{center}
\caption{\label{tab:syst_MC} Summary of the various sources of uncertainty for the non-multi-jet background estimates in the signal
  regions. The first row shows the uncertainty due to the limited statistics in the MC samples. The second row shows the total systematic
  uncertainty from theory. The main experimental systematic uncertainties are given in the rows labelled with ``Tau ID and trigger'' and ``Tau
  Energy scale'', while the row ``Others'' shows the contribution from the remaining sources of experimental systematic uncertainty as
  described in the text.
  The ``Total'' uncertainty is the sum in quadrature of each source.
} \smallskip
\begin{tabular}{c|c|c|c|c}
\hline
Source                                & $W$+jets      & Diboson  & $Z$+jets & Top  \\\hline
MC statistics                                         &     16--36\%    & 15--28\%  & 44--80\%       & 23--50\%   \\
Theoretical uncertainty                          &    17--30\%     & 17--27\%  & 25--30\%       & 10--20\%   \\
Tau ID and trigger                                  &     10--18\%&  20--21\%  &  10--20\%       &  22--28\%      \\ 
Tau Energy Scale                                    &     12--20\%    &3--13\%    &  4--12\%       &   2--7\%      \\
Others                                                    &      1--10\%    &3--9\%    &  5--10\%       &   10--20\%      \\

\hline     
Total                                                     &     34--48\%       &  35--44\%  &  58--85\%       &  43--62\%      \\
\hline     
\end{tabular}
\end{center}
\end{table}

Signal cross sections are calculated to NLO.  
Their uncertainties are taken from an envelope of cross-section predictions 
using different PDF sets and factorisation and renormalisation scales
\cite{Kramer:2012bx}. 
Uncertainties associated with modelling of ISR in SUSY signal samples were evaluated for a few benchmark samples and found to be negligible.
Systematic uncertainties associated with the signal
selection efficiency include those due to the tau trigger efficiency, tau identification and energy scale, jet reconstruction and \met\
calculation. The uncertainty on the integrated luminosity also affects the predicted signal yield. The total uncertainty varies between 20\% and 30\% for SUSY scenarios 
to which this measurement is sensitive. The dominant experimental
uncertainty is due to the tau trigger efficiency, and is around 20\%.

\section{Results}
\label{sec:result}

The observed number of events in each signal region and the expected contributions from SM processes  
are given in table \ref{tab:results}. The contributions of multi-jet and $W$+jets events were scaled with the normalisation factors
obtained from the fit described in section~\ref{sec:BG_fit}. The
normalisation factors do not deviate from unity by more than 1\%. Due to the limited number of events in the multi-jet CRs, the
uncertainty on the multi-jet normalisation varies between 54\% and 131\% depending on the CR, while the uncertainty on the $W$+jets normalisation is around 10\%.
All statistical uncertainties arising from the limited number of MC events are included. The effect of limited data events in the CRs is included in the systematic uncertainty. All systematic uncertainties except for the jet energy scale and jet energy resolution are obtained taking into account the correlations among control regions and background processes. 
The total uncertainties are the sum in quadrature of systematic and statistical uncertainties. 
Due to the simultaneous fit, the uncertainty on the $W$+jets
estimation decreases by 10\% with respect to the numbers in table
\ref{tab:syst_MC}, while the uncertainties on the remaining
non-multi-jet backgrounds vary by less than 2\% depending on the SR.
Due to the large uncertainty on the multi-jet normalisation, the
uncertainty on the multi-jet estimate increases by up to 25\% depending on the SR with respect to the numbers quoted in table \ref{tab:qcd-est-result}. 
Agreement is found between observations and background expectations within $1\sigma$ in all signal regions except for SR-C1C1. 
In SR-C1C1, the observed number of events fluctuates below the number
of expected SM events by $2.2\sigma$, mainly in the region 250~GeV~$< m_{{\rm T}\tau1}+m_{{\rm T}\tau2}<$~270~GeV.
 
For each SR, the significance of a possible excess of observed events over the SM prediction 
is quantified by the one-sided probability, $p_0$, of the background alone to fluctuate to the observed number of events or higher
using the asymptotic formula described in~\cite{statforumlimitsl}. A fit similar to the one described in section~\ref{sec:BG_fit} is used, except that the
number of events observed in the SR is added as an input to the fit, and an additional parameter for the non-SM signal strength,
constrained to be non-negative, is fit.
Upper limits at 95\% confidence level (CL) on the number of non-SM events in the SRs are derived using the CLs prescription ~\cite{CLs:2002} and
neglecting any possible signal contamination in the control regions. Normalising these by the integrated luminosity of the data
sample, they can be interpreted as upper limits on the visible non-SM cross section, $\sigma_{\rm {vis}}^{95}$, which is
defined as the product of acceptance, reconstruction efficiency and
production cross section. All systematic uncertainties and their
correlations are taken into account via nuisance parameters.
The accuracy of the limits obtained by the asymptotic formula was tested for all SRs by randomly generating a large number of pseudo-datasets and repeating the fit, and good agreement was found.
 The results are given in table~\ref{tab:results}. 

\begin{table}
\centering
\caption{
 Observed and expected numbers of events  in the signal regions for 20.3~fb$^{-1}$. 
The contributions of multi-jet and $W$+jets events were scaled with the normalisation factors
obtained from the fit described in section~\ref{sec:BG_fit}. 
The shown uncertainties are the sum in quadrature of statistical  and systematic uncertainties. The correlation of systematic uncertainties among control 
regions and background processes is fully taken into account and, as a result, the numbers given here may be different from those in tables \ref{tab:qcd-est-result} and \ref{tab:syst_MC}.
Expected event yields for the SUSY reference points (see
section \ref{sec:signalGrids}) are also shown.
The one-sided $p_0$-values and the observed and expected 95\%
CL upper limits on the visible non-SM cross section (\limsigvis), 
obtained from the fit described in section~\ref{sec:result}, are given. Values of $p_0>0.5$  are truncated to $p_0=0.5$.
 }
 \smallskip
\begin{tabular}{|c|c|c|c|c|}
\hline
SM process                      & SR-C1N2                           & SR-C1C1                & SR-DS-highMass    & SR-DS-lowMass\\
\hline
Top                 &  0.30 $\pm$ 0.19           & 0.7 $\pm$ 0.4        & 0.9 $\pm$ 0.4   & 1.3 $\pm$ 0.6 \\
$Z$+jets        &  0.9 $\pm$ 0.5         & 0.20 $\pm$ 0.17        & 0.6 $\pm$ 0.4   & 0.40 $\pm$ 0.27 \\
$W$+jets       &  2.2 $\pm$ 0.8          & 11.2 $\pm$ 2.8      & 2.7 $\pm$ 0.9   & 4.1 $\pm$ 1.2 \\
Diboson         &  2.2 $\pm$ 0.9            & 3.8 $\pm$ 1.4        & 2.5 $\pm$ 1.0   & 2.9 $\pm$ 1.0 \\
Multi-jet        &   2.3 $\pm$ 2.0          & 5.8 $\pm$ 3.3        & 0.9 $\pm$ 1.2   & 2.8 $\pm$ 2.3 \\
\hline
SM total            &     7.9 $\pm$ 2.4             & 22 $\pm$ 5           & 7.5 $\pm$ 1.9       &   11.5 $\pm$ 2.9\\
\hline
Observed           &     11                           & 12                    &  7              &  15  \\
\hline     
\hline     
Ref. point 1             & 11.3 $\pm$ 2.8        & 8.5 $\pm$ 2.2   &   10.2  $\pm$ 2.6  & 7.5 $\pm$ 2.0 \\
Ref. point 2             & 9.2 $\pm$ 2.1            &20 $\pm$ 4     &   12.4 $\pm$ 2.8   & 12.8 $\pm$ 2.7 \\ 
Ref. point 3             & 0.8  $\pm$ 0.5        &7.6  $\pm$ 1.9 &   3.8 $\pm$ 1.0 & 5.2 $\pm$ 1.3\\
\hline   
\hline     
$p_0$                                        &  0.20                &0.50               & 0.50       & 0.21 \\
Expected $\sigma_{\rm {vis}}^{95}$ (fb)   & $<{0.42}^{+0.19}_{-0.11}$ & $<{0.56}^{+0.25}_{-0.14}$ & $<{0.37}^{+0.17}_{-0.10}$ & $<{0.51}^{+0.18}_{-0.15}$ \\
Observed $\sigma_{\rm {vis}}^{95}$ (fb) &     $<0.59$                         & $<0.37$           &  $<0.37$            &  $<0.66$ \\

\hline
\end{tabular}
\label{tab:results}
\end{table}

In figure \ref{fig:SR-result} the distributions of the relevant kinematic variables are shown for data, SM expectations and illustrative SUSY benchmark models.
The SM background distributions are taken from MC simulation, except for the multi-jet contribution, which is estimated using the ABCD method described in section
\ref{sec:BG_QCD}. The normalisation factors obtained from the fit detailed in section \ref{sec:BG_fit} are used to correct the 
expected distributions of the $W$+jets and multi-jet processes.

\begin{figure}
\centering
\subfigure[~SR-C1N2]{  \includegraphics[width=0.45\textwidth]{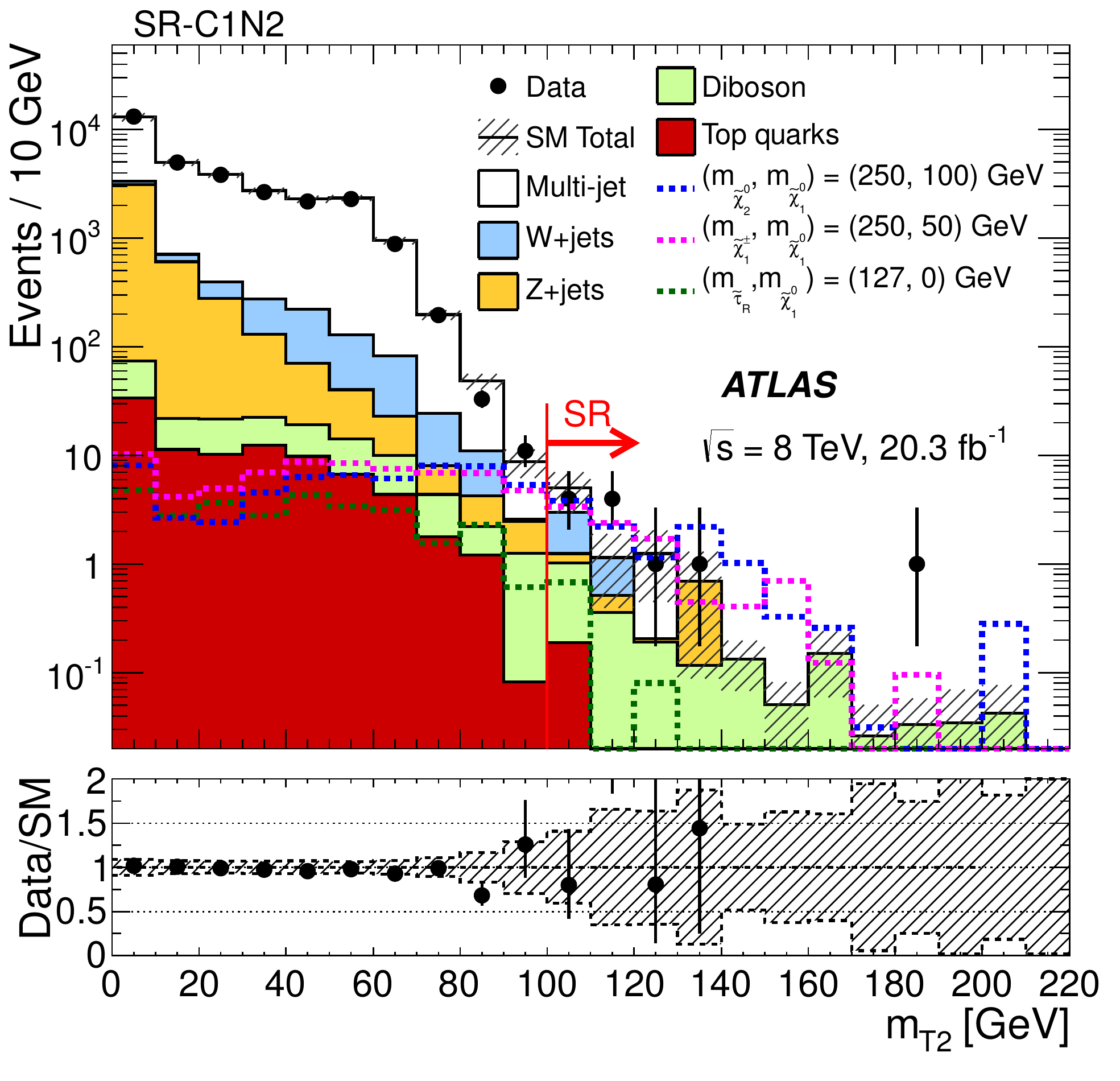} }
\subfigure[~SR-C1C1]{   \includegraphics[width=0.45\textwidth]{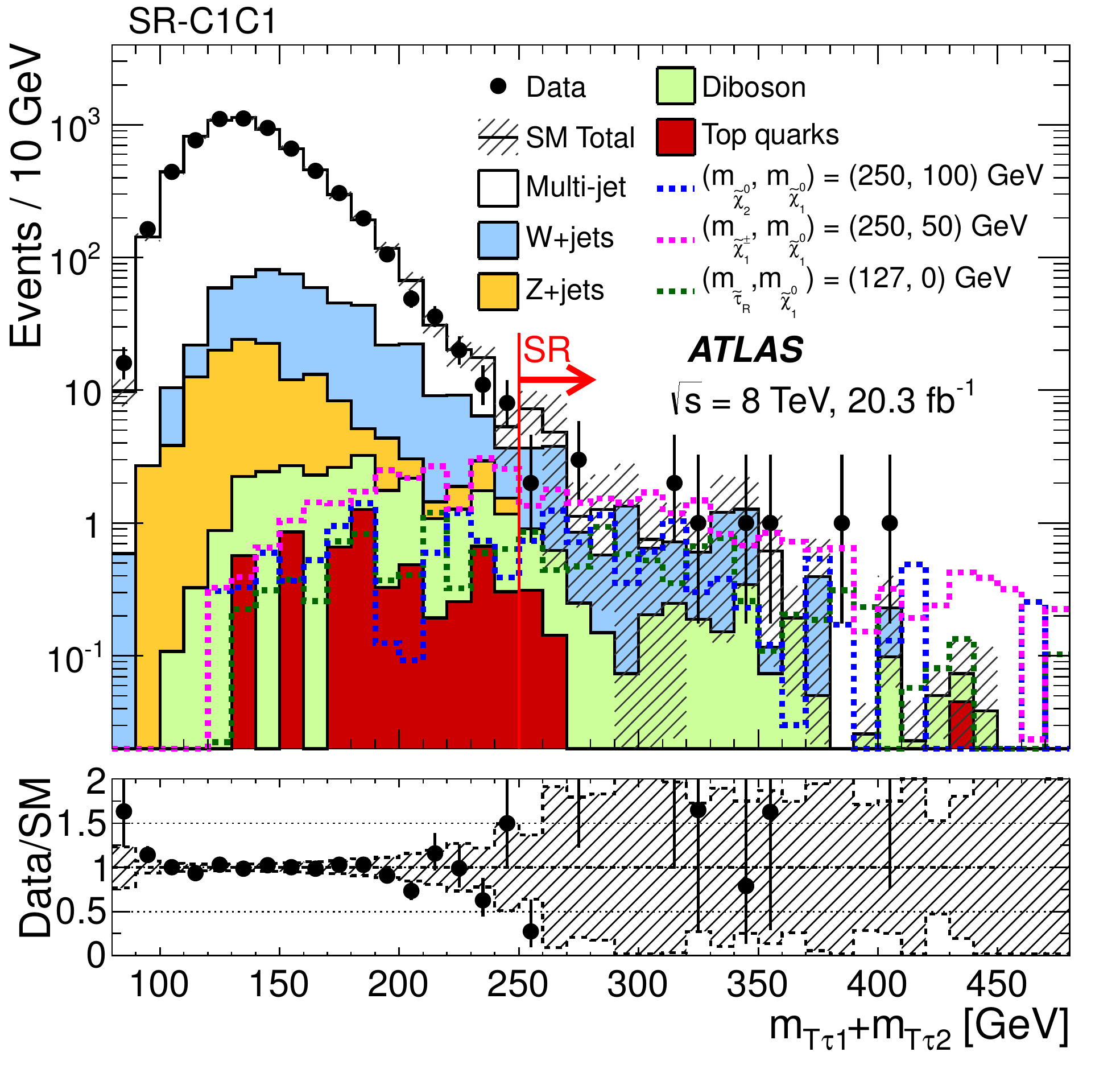}  }\\
\subfigure[~SR-DS-highMass]{   \includegraphics[width=0.45\textwidth]{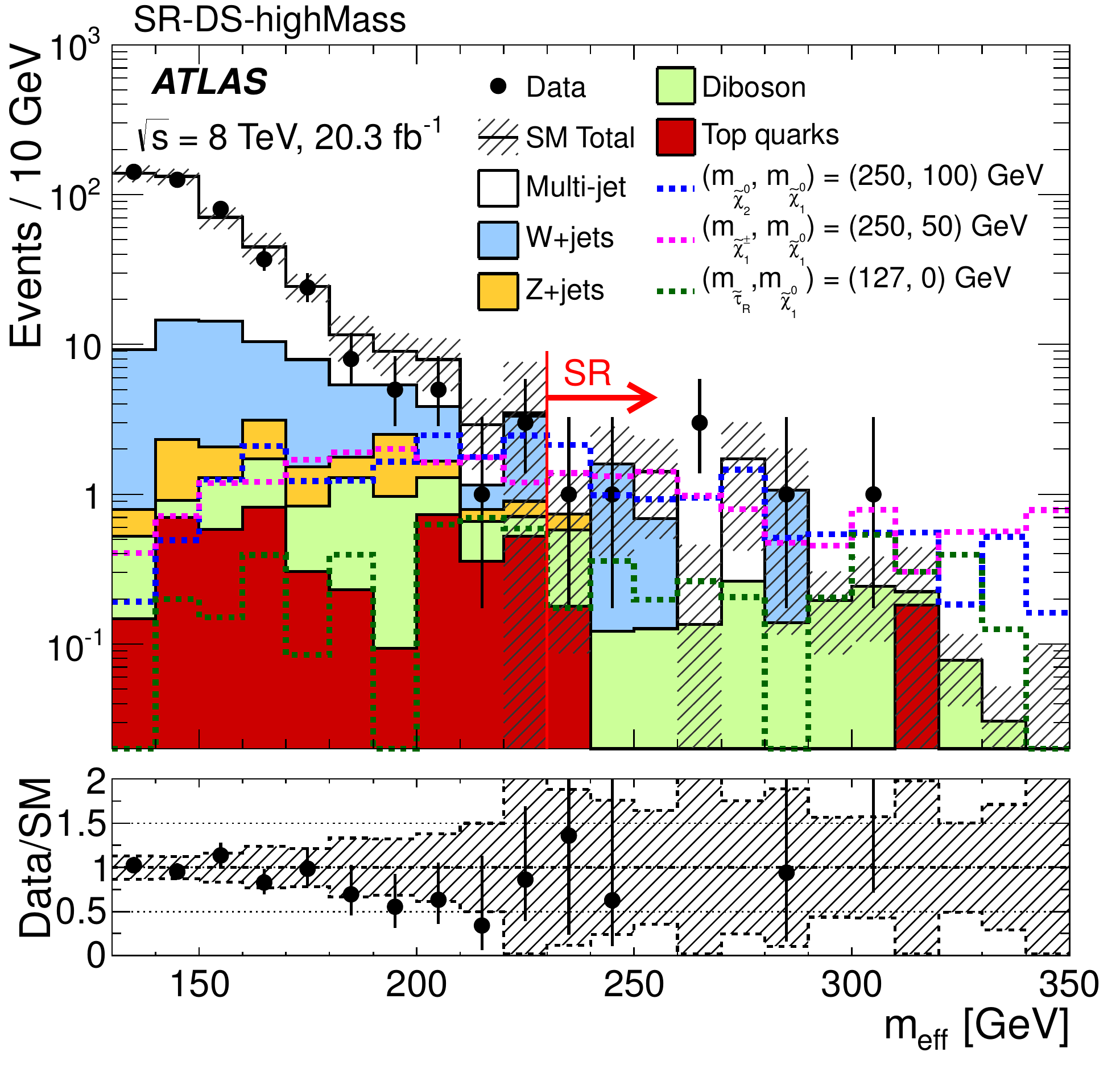}  }
\subfigure[~SR-DS-lowMass]{   \includegraphics[width=0.45\textwidth]{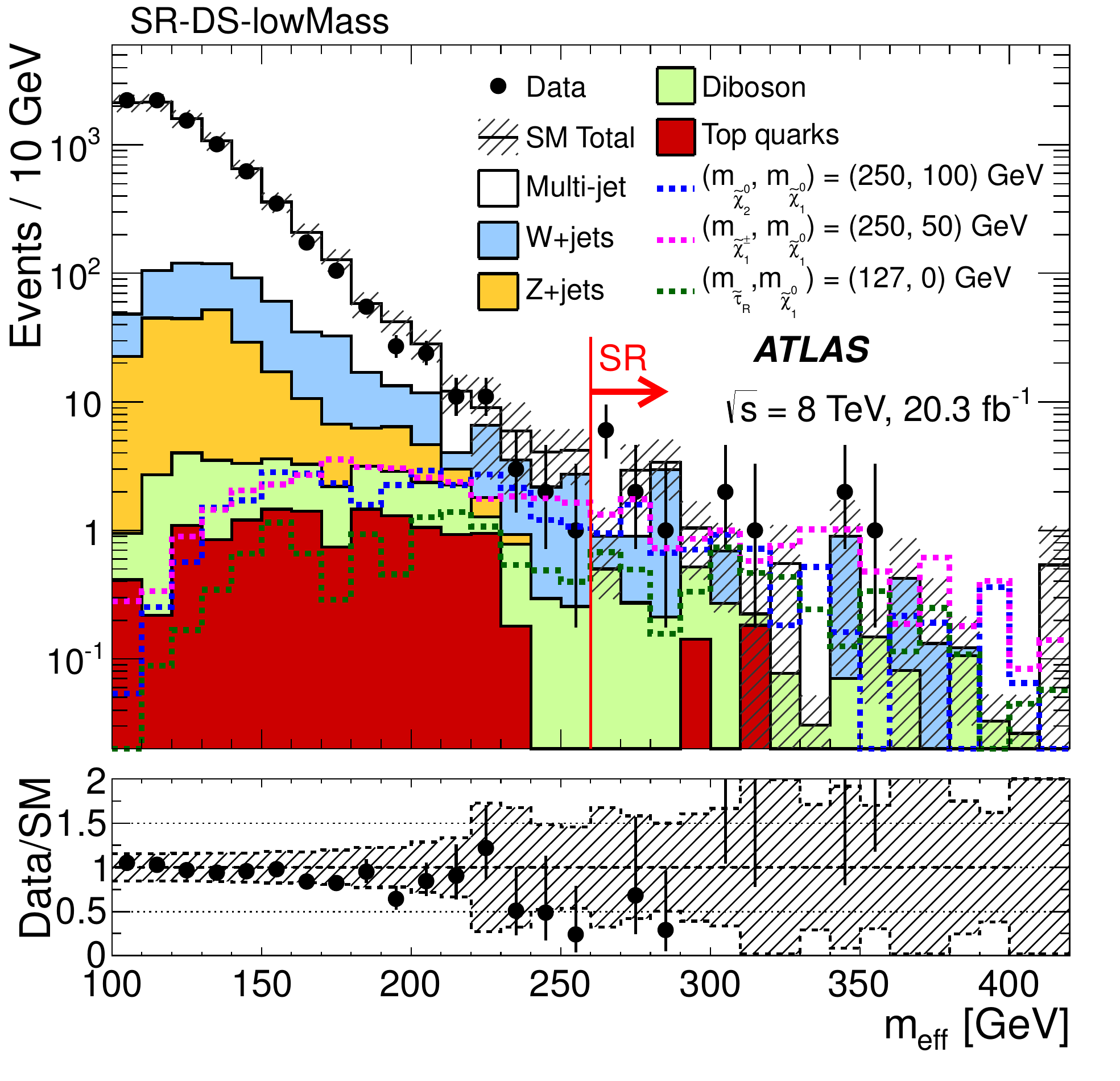}  }
\caption{Distributions of relevant kinematic variables before the requirement
   on the given variable is applied: (a) \mttwo~for SR-C1N2, (b) $m_{{\rm T}\tau1}+m_{{\rm T}\tau2}$ for
 SR-C1C1, (c) $m_{\rm eff}$ for SR-DS-highMass, and
(d) $m_{\rm eff}$ for SR-DS-lowMass. 
The stacked histograms show the expected SM backgrounds normalised to 20.3~\ifb.
The multi-jet contribution is estimated from data using the ABCD method. 
The hatched bands represent the sum in quadrature of systematic and
statistical uncertainties on the total SM background.
The lower panels show the ratio of data to the total SM background estimate. 
}
\label{fig:SR-result}
\end{figure}

\section{Interpretation}
\label{section:interpretation}

In the absence of a significant excess over the SM background
expectations, the observed numbers of events in the signal regions are used to
place model-dependent exclusion limits at 95\% CL for 
the pMSSM and the simplified models described in section~\ref{sec:signalGrids}.
The same CLs limit-setting procedure as described in section \ref{sec:result} is used,
except that the SUSY signal is allowed to populate both the signal and the control regions.
All SRs defined in section \ref{sec:SR}  are considered in order to derive limits for
all the SUSY models considered in this paper, regardless of the
specific production mode for which they were optimised. Since the SRs are not mutually
exclusive, for each point in the parameter space the SR which gives the best expected limit is used.

The results are shown in figures \ref{fig:alimit}--\ref{fig:1DlimitDS}. The solid (dashed) lines show the observed (expected) 
exclusion contours. The band around the expected limit shows the
$\pm1\sigma$ variations, including all
uncertainties except theoretical uncertainties on the signal cross section. 
The dotted lines around the observed limit indicate the sensitivity to $\pm1\sigma$ variations of the theoretical uncertainties. 
All mass limits hereafter quoted correspond to the observed limits
reduced by $1\sigma$ of the signal cross sections.

\subsection{Simplified models: chargino--neutralino and chargino--chargino production}
The exclusion limits for the simplified models
characterised by $\tilde{\chi}_{1}^{\pm}$$\tilde{\chi}_{2}^{0}$ and
$\tilde{\chi}_{1}^{\pm}$$\tilde{\chi}_{1}^{\mp}$ production
with intermediate staus are shown in figure \ref{fig:alimit}. 
In figure~\ref{fig:aSRbest}, both production processes are considered simultaneously, whereas in figure~\ref{fig:cSRbest}
only chargino--chargino production is assumed.

Chargino masses up to 345 GeV are excluded for a massless lightest neutralino in the scenario of direct production of chargino pairs.
In the case of associated production of mass-degenerate charginos and next-to-lightest neutralinos, chargino masses up to 410 GeV are
excluded for a massless lightest neutralino. These limits improve the results from the ATLAS three-lepton analysis~\cite{3Lep-2012}, where only production
of $\tilde{\chi}_{1}^{\pm}$$\tilde{\chi}_{2}^{0}$ was considered due
to the low sensitivity to the decays of $\tilde{\chi}_{1}^{\pm}$$\tilde{\chi}_{1}^{\mp}$.
For the scenario of direct production of chargino pairs, SR-C1C1
(SR-DS-highMass) provides the best exclusion limit for low (high) chargino masses, while for the associated production of mass-degenerate
charginos and next-to-lightest neutralinos, SR-C1N2 and SR-DS-highMass have the highest sensitivity over the whole parameter space.

\begin{figure}
  \centering
  \subfigure[~$\tilde{\chi}_{1}^{\pm}\tilde{\chi}_{1}^{\mp}$ and $\tilde{\chi}_{1}^{\pm}\tilde{\chi}_{2}^{0}$ production]{
    \includegraphics[width=0.475\textwidth]{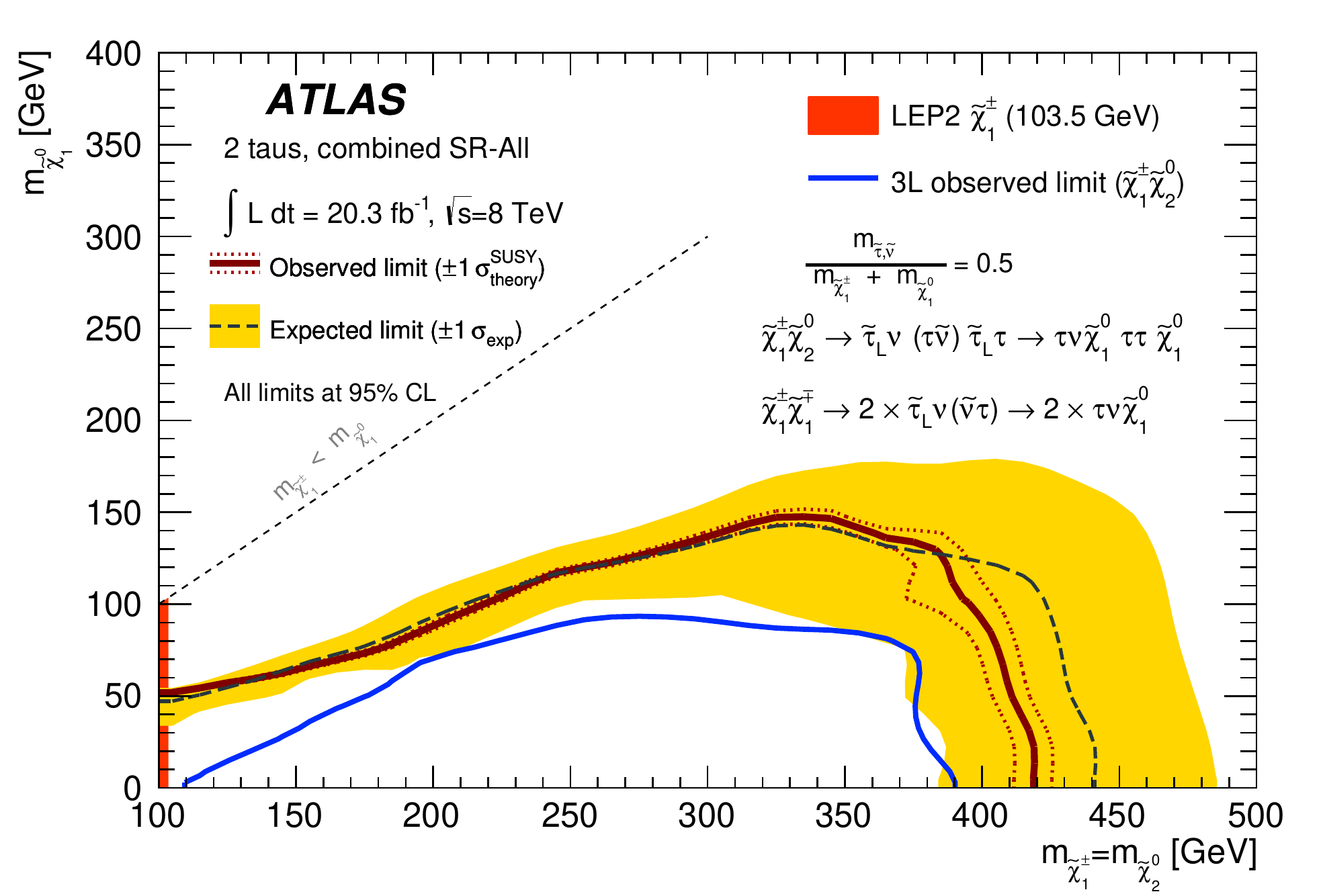}
    \label{fig:aSRbest}
  }
 \subfigure[~$\tilde{\chi}_{1}^{\pm}\tilde{\chi}_{1}^{\mp}$ production]{
    \includegraphics[width=0.475\textwidth]{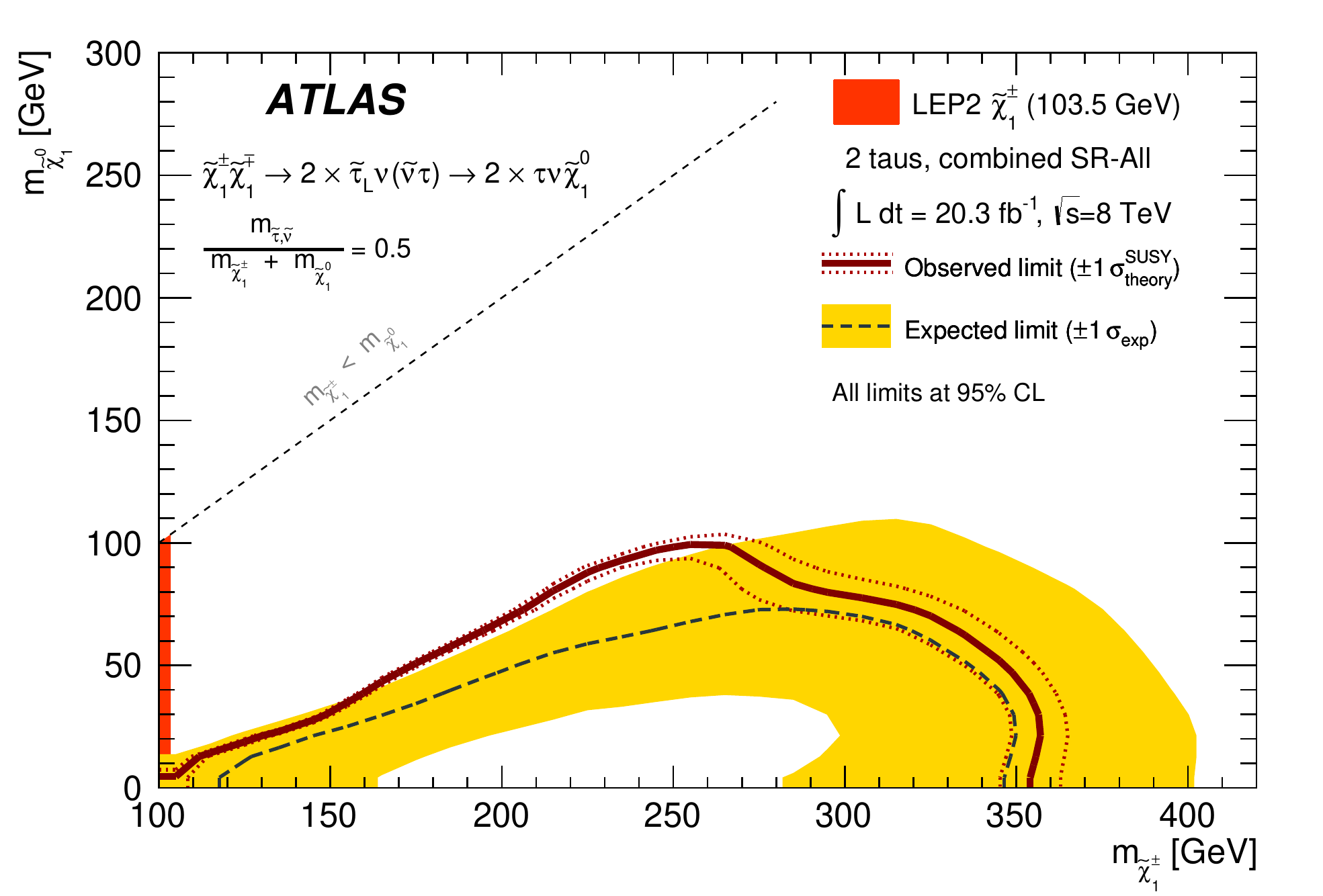}
    \label{fig:cSRbest}    
  }

  \caption{95\% CL exclusion limits for simplified models with (a) a
    combination of chargino--neutralino and chargino--chargino
  production  and (b) chargino--chargino production only. 
See text for details of exclusion curves and uncertainty bands.
Also shown is the LEP limit \cite{lepsusy} on the mass of the chargino.
The blue contour in (a) corresponds to the observed limit from the ATLAS three-lepton analysis~\cite{3Lep-2012}, where only $\tilde{\chi}_{1}^{\pm}$$\tilde{\chi}_{2}^{0}$ production
was considered.
} 
  \label{fig:alimit}
\end{figure}

\subsection{Direct stau production}

Due to the low cross section, the sensitivity of this analysis to the direct production of stau pairs is degraded relative to the sensitivity obtained for models of electroweak production of charginos and neutralinos.
Upper limits on the cross section were derived for the direct stau production model described in section \ref{sec:signalGrids}, and are shown
separately for the production of $\tilde{\tau}_{\rm R} \tilde{\tau}_{\rm R}$ and $\tilde{\tau}_{\rm L} \tilde{\tau}_{\rm L}$ in figure \ref{fig:xsectionDS}. 
For large stau masses, SR-DS-highMass provides the best upper limits,
while SR-C1C1 has the best performance for low stau masses. 
The results in SR-C1C1 lead to a stronger observed exclusion limit
than expected due to the number of observed events being fewer than the SM prediction.

The signal strength is defined as the scaling factor that should be
applied to the theoretical cross section to exclude the considered model at 95\% CL. 
The upper limit on the signal strength for the associated production of $\tilde{\tau}_{\rm R} \tilde{\tau}_{\rm R}$ and
$\tilde{\tau}_{\rm L} \tilde{\tau}_{\rm L}$, for different lightest neutralino masses and
as a function of the $\tilde{\tau}_{\rm R}$ mass is shown in figure
\ref{fig:1DlimitDS}. 

 The best observed upper limit on the signal strength is found for a mass of the $\tilde{\tau}_{\rm R}$ ($\tilde{\tau}_{\rm L}$) of 
90.6 (93.1) GeV and~a massless~$\tilde{\chi}_{1}^{0}$. For this combination of stau and LSP masses, the theoretical cross section at NLO is 
0.07 (0.17) pb for $\tilde{\tau}_{\rm R} \tilde{\tau}_{\rm R}$ ($\tilde{\tau}_{\rm L} \tilde{\tau}_{\rm L}$) production, while the excluded cross section is 0.22 (0.28) pb and 
the upper limit on the signal strength for the combined production of
$\tilde{\tau}_{\rm R} \tilde{\tau}_{\rm R}$ and $\tilde{\tau}_{\rm L} \tilde{\tau}_{\rm L}$ is 0.95.

 \begin{figure}
  \centering
\subfigure[]{
      \includegraphics[width=0.475\textwidth]{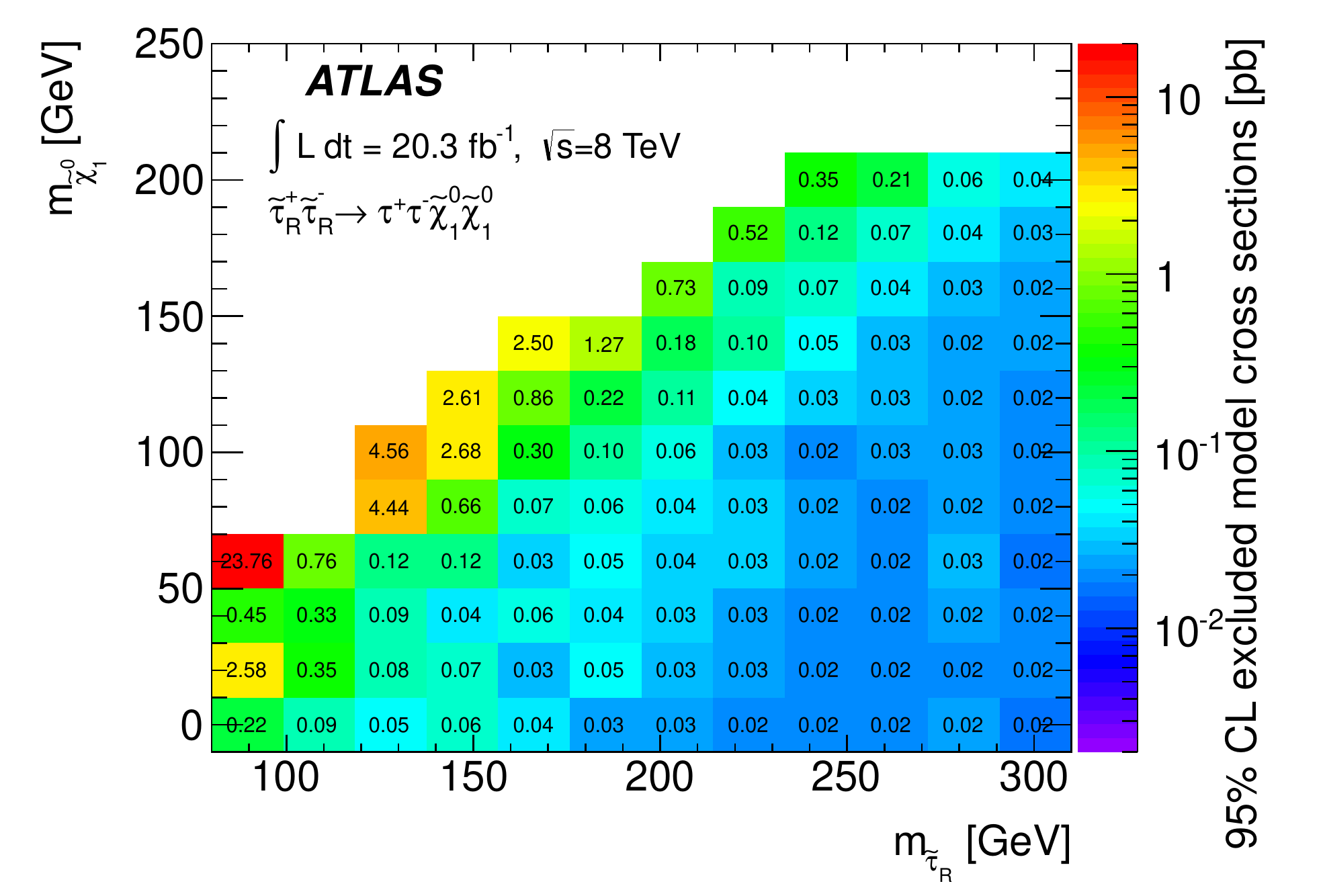}
}
\subfigure[]{
   \includegraphics[width=0.475\textwidth]{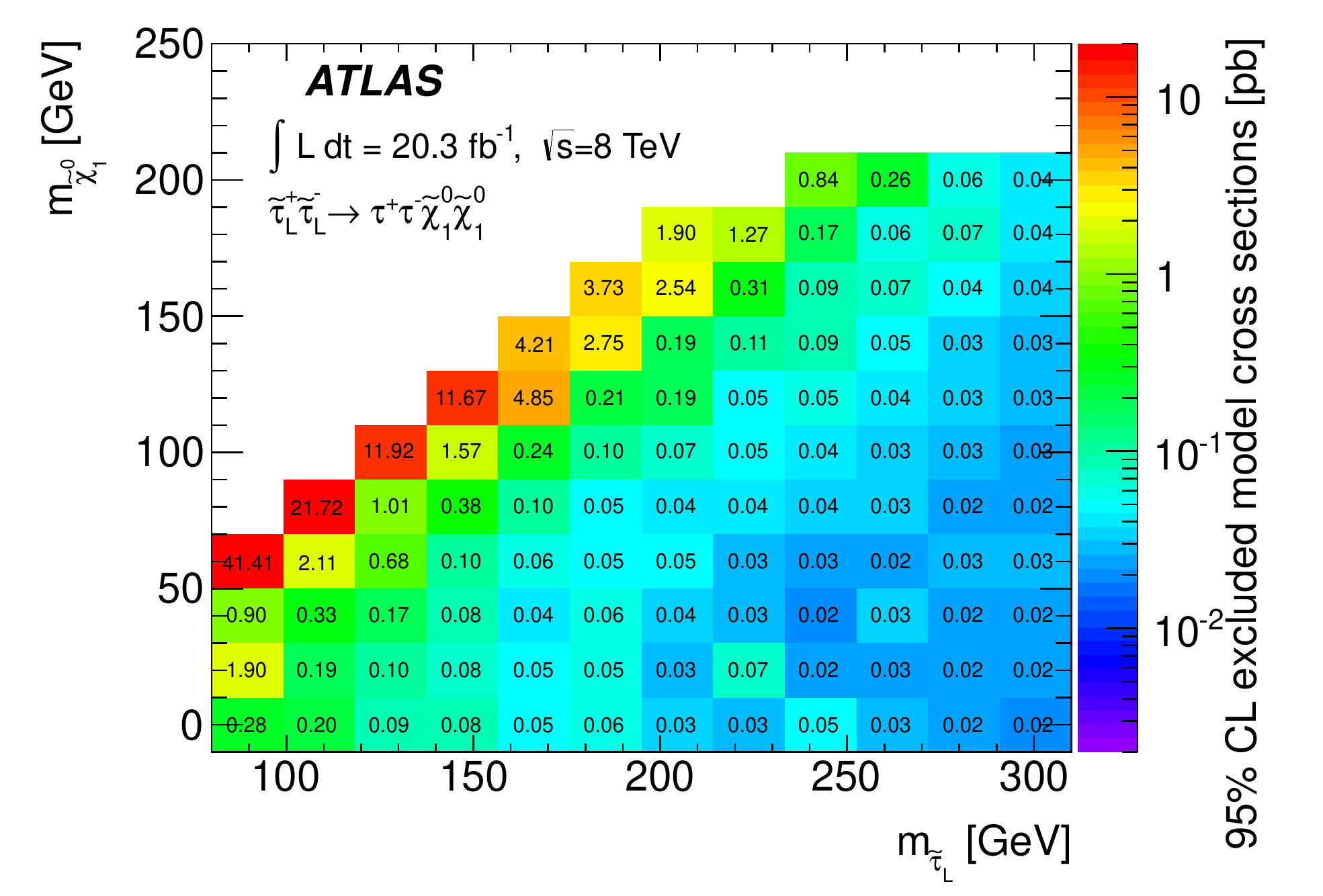}
}
\caption{Upper limits on the cross section for production 
of only (a) $\tilde{\tau}_{\rm R} \tilde{\tau}_{\rm R}$ or (b) $\tilde{\tau}_{\rm L} \tilde{\tau}_{\rm L}$ pairs.
}       
\label{fig:xsectionDS}
\end{figure}

 \begin{figure}
  \centering
    \includegraphics[width=1.\textwidth]{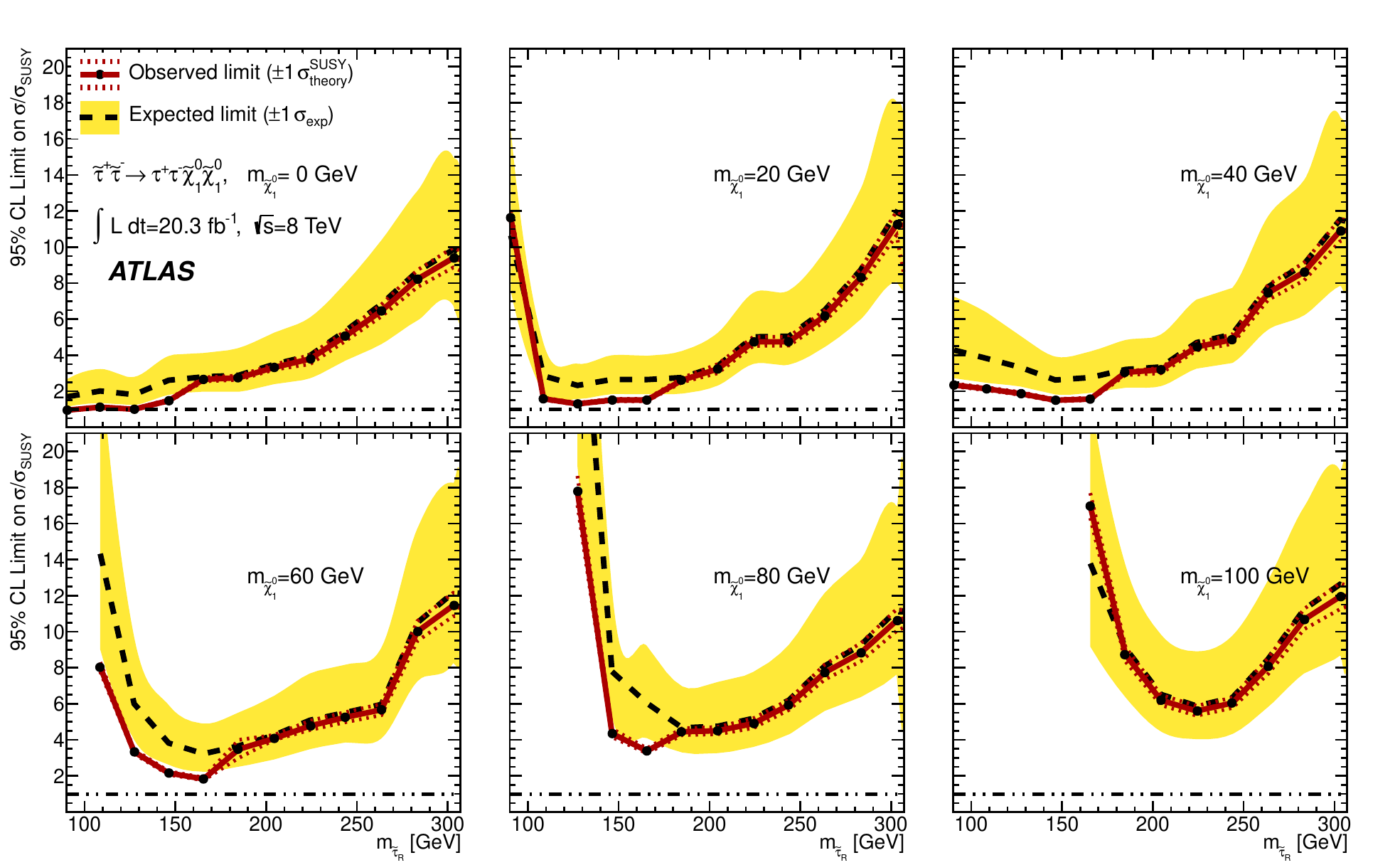}
\caption{Upper limit on the signal strength for the associated production of $\tilde{\tau}_{\rm R} \tilde{\tau}_{\rm R}$ and
$\tilde{\tau}_{\rm L} \tilde{\tau}_{\rm L}$, for different lightest neutralino masses and as a function of the $\tilde{\tau}_{\rm R}$ mass. 
See text for details of exclusion curves and uncertainty bands. }       
\label{fig:1DlimitDS}
\end{figure}

\subsection{The pMSSM model}

Limits on the mass parameters $M_2$ and $\mu$ are set within the pMSSM framework with parameters described in section \ref{sec:signalGrids}. 
In figure~\ref{fig:pmssmFixed}, the exclusion limits for the pMSSM model with fixed stau mass are shown in the $\mu$-$M_{2}$ plane.
The region at low $M_2$ cannot be excluded since it corresponds to points in the parameter space 
where the chargino and neutralino are lighter than the stau.
Since in this model the cross section for direct stau production is constant, this process dominates in the remaining allowed region at large $M_2$ and 
$\mu$. Direct stau production accounts for 60\% of the events for $M_2 = \mu = 400$ GeV, and 85\% of the events for $M_2 = \mu = 500$ GeV.

 Figure~\ref{fig:pmssmVariable} shows the exclusion limits in the $\mu$-$M_{2}$ plane for the pMSSM model with variable stau mass.  
For both pMSSM models, the excluded $\tilde{\chi}_{1}^{\pm}$
($\tilde{\chi}_{2}^{0}$) mass range is 100--350 GeV as can be seen from the light grey iso-mass lines
of $\tilde{\chi}_{1}^{\pm}$ ($\tilde{\chi}_{2}^{0}$). For values of $\mu$ larger than those simulated for this analysis, 
the pMSSM phenomenology is similar to that studied here. For larger values of $M_2$, the production cross section of heavier neutralinos and 
charginos increases. In general, the shown limits on the lightest chargino mass can be expected to be similar also at large $\mu$ ($M_2$) 
for values of $M_2$ ($\mu$) in the range 150--350 (100--300) GeV.

In the pMSSM model with fixed stau mass, SR-DS-highMass provides better
exclusion at high $\mu$, $M_2$. For $M_2, \mu < 200$ GeV, SR-C1N2 and
SR-C1C1 provide the most stringent limits. 
In the pMSSM model with variable stau mass, SR-C1N2 and SR-DS-highMass give the best sensitivity in the whole parameter space. 

\begin{figure}
  \centering
    \subfigure[~pMSSM model (fixed stau mass) ]{  \includegraphics[width=0.475\textwidth]{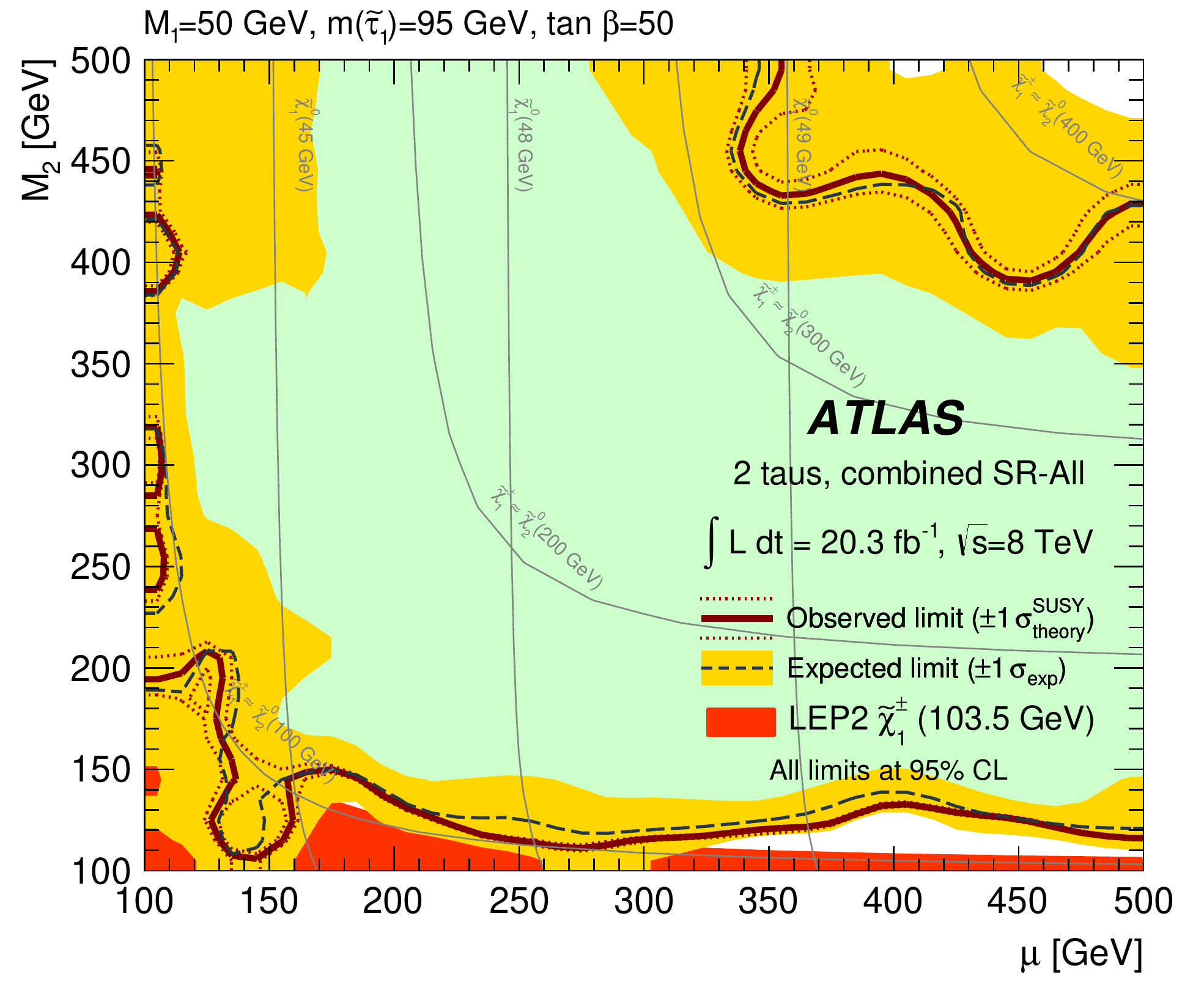} \label{fig:pmssmFixed}}
    \subfigure[~pMSSM model (variable stau mass) ]{  \includegraphics[width=0.475\textwidth]{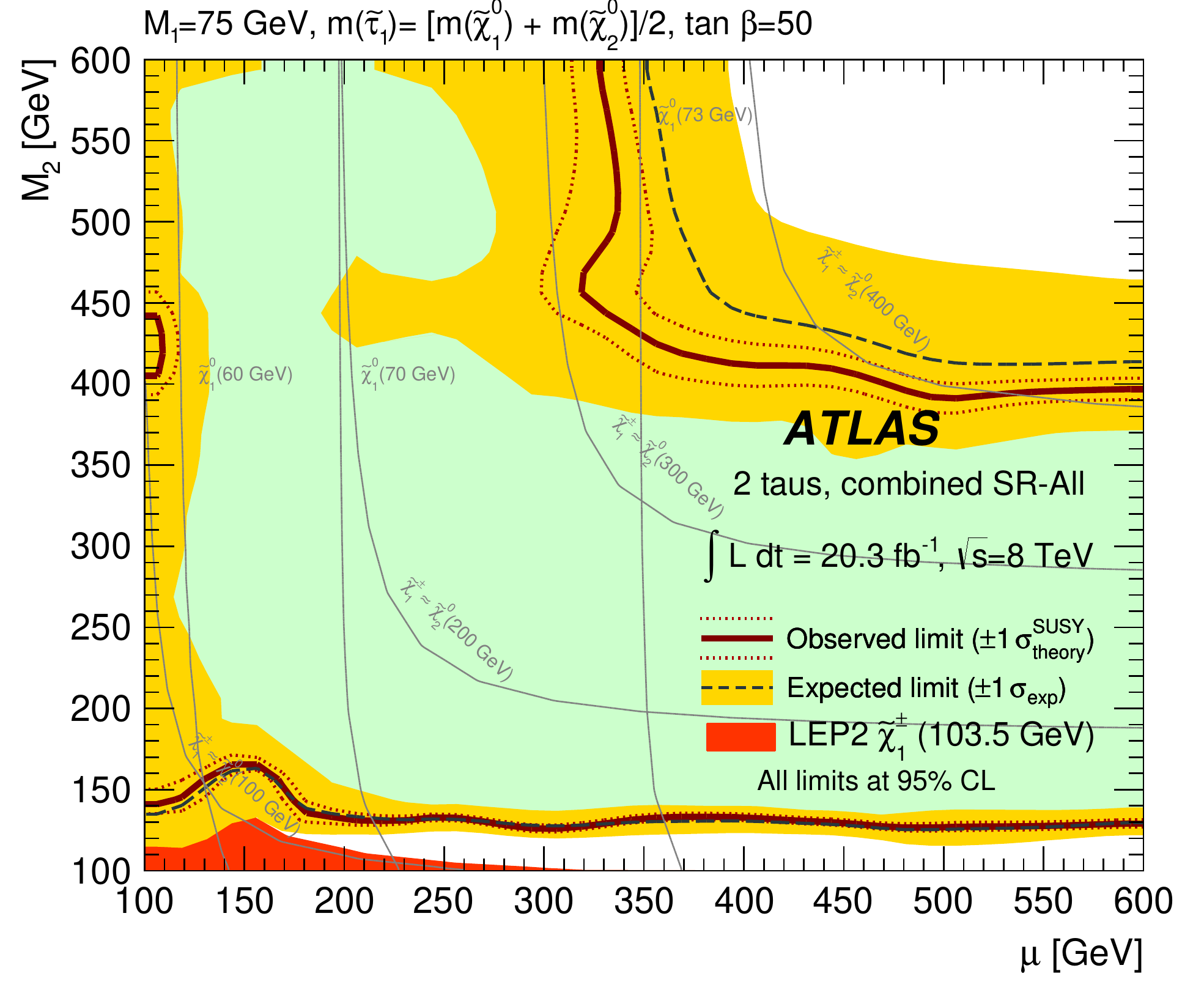} \label{fig:pmssmVariable}}
 \caption{95\% CL exclusion limits in the $\mu$--$M_{2}$ mass plane for
   the pMSSM models with (a) fixed and (b) variable stau mass. See
   text for details of exclusion curves and uncertainty bands.
The areas excluded by the $-1\sigma$ expected limit are shown in green.
The LEP limit \cite{lepsusy} on the mass of the chargino is also shown in red.
}
  \label{fig:pmssmlimit}
\end{figure}

\section{Conclusion}
 
Searches for the electroweak production of supersymmetric particles in events with
at least two hadronically decaying taus, missing transverse
momentum and low jet activity in the final state are performed using 20.3~fb$^{-1}$ of
proton--proton collision data at $\sqrt{s}=8$~TeV recorded with the ATLAS experiment at the Large Hadron Collider.
Agreement between data and SM expectations is observed in all signal regions.
These results are used to set limits on the visible cross section
for non-SM events in each signal region. Exclusion limits
are placed on parameters of the pMSSM and simplified models.

For simplified models, chargino masses up to 345 GeV are excluded for a massless lightest
neutralino in the scenario of direct production of wino-like
chargino pairs, with each chargino decaying into the lightest neutralino via an intermediate on-shell stau. 
In the case of associated production of mass-degenerate charginos and next-to-lightest neutralinos, masses up to 410 GeV are
excluded for a massless lightest neutralino.
In pMSSM models the excluded $\tilde{\chi}_{1}^{\pm}$
($\tilde{\chi}_{2}^{0}$) mass range is between 100 and 350 GeV.
For direct stau production, the best upper limit on the signal strength
is found for a  mass of the $\tilde{\tau}_{\rm R}$ ($\tilde{\tau}_{\rm L}$) of 90.6 (93.1)~GeV and a massless~$\tilde{\chi}_{1}^{0}$. 
The excluded cross section for $\tilde{\tau}_{\rm R} \tilde{\tau}_{\rm R}$ ($\tilde{\tau}_{\rm L} \tilde{\tau}_{\rm L}$) is 0.22 (0.28) pb for this combination of stau and lightest supersymmetric particle masses, while 
the theoretical cross section at NLO is 0.07 (0.17) pb.




\section{Acknowledgements}

We thank CERN for the very successful operation of the LHC, as well as the
support staff from our institutions without whom ATLAS could not be
operated efficiently.

We acknowledge the support of ANPCyT, Argentina; YerPhI, Armenia; ARC,
Australia; BMWFW and FWF, Austria; ANAS, Azerbaijan; SSTC, Belarus; CNPq and FAPESP,
Brazil; NSERC, NRC and CFI, Canada; CERN; CONICYT, Chile; CAS, MOST and NSFC,
China; COLCIENCIAS, Colombia; MSMT CR, MPO CR and VSC CR, Czech Republic;
DNRF, DNSRC and Lundbeck Foundation, Denmark; EPLANET, ERC and NSRF, European Union;
IN2P3-CNRS, CEA-DSM/IRFU, France; GNSF, Georgia; BMBF, DFG, HGF, MPG and AvH
Foundation, Germany; GSRT and NSRF, Greece; ISF, MINERVA, GIF, I-CORE and Benoziyo Center,
Israel; INFN, Italy; MEXT and JSPS, Japan; CNRST, Morocco; FOM and NWO,
Netherlands; BRF and RCN, Norway; MNiSW and NCN, Poland; GRICES and FCT, Portugal; MNE/IFA, Romania; MES of Russia and ROSATOM, Russian Federation; JINR; MSTD,
Serbia; MSSR, Slovakia; ARRS and MIZ\v{S}, Slovenia; DST/NRF, South Africa;
MINECO, Spain; SRC and Wallenberg Foundation, Sweden; SER, SNSF and Cantons of
Bern and Geneva, Switzerland; NSC, Taiwan; TAEK, Turkey; STFC, the Royal
Society and Leverhulme Trust, United Kingdom; DOE and NSF, United States of
America.

The crucial computing support from all WLCG partners is acknowledged
gratefully, in particular from CERN and the ATLAS Tier-1 facilities at
TRIUMF (Canada), NDGF (Denmark, Norway, Sweden), CC-IN2P3 (France),
KIT/GridKA (Germany), INFN-CNAF (Italy), NL-T1 (Netherlands), PIC (Spain),
ASGC (Taiwan), RAL (UK) and BNL (USA) and in the Tier-2 facilities
worldwide.

\FloatBarrier

\bibliographystyle{JHEP}
\bibliography{DG2Tau}



\onecolumn 
\clearpage 
\begin{flushleft}
{\Large The ATLAS Collaboration}

\bigskip

G.~Aad$^{\rm 84}$,
B.~Abbott$^{\rm 112}$,
J.~Abdallah$^{\rm 152}$,
S.~Abdel~Khalek$^{\rm 116}$,
O.~Abdinov$^{\rm 11}$,
R.~Aben$^{\rm 106}$,
B.~Abi$^{\rm 113}$,
M.~Abolins$^{\rm 89}$,
O.S.~AbouZeid$^{\rm 159}$,
H.~Abramowicz$^{\rm 154}$
% ...
\bigskip
\\
% ...
$^{11}$ Institute of Physics, Azerbaijan Academy of Sciences, Baku, Azerbaijan\\
% ...
$^{84}$ CPPM, Aix-Marseille Universit{\'e} and CNRS/IN2P3, Marseille, France\\
% ...
$^{89}$ Department of Physics and Astronomy, Michigan State University, East Lansing MI, United States of America\\
% ...
$^{106}$ Nikhef National Institute for Subatomic Physics and University of Amsterdam, Amsterdam, Netherlands\\
% ...
$^{112}$ Homer L. Dodge Department of Physics and Astronomy, University of Oklahoma, Norman OK, United States of America\\
$^{113}$ Department of Physics, Oklahoma State University, Stillwater OK, United States of America\\
% ...
$^{116}$ LAL, Universit{\'e} Paris-Sud and CNRS/IN2P3, Orsay, France\\
% ...
$^{152}$ Institute of Physics, Academia Sinica, Taipei, Taiwan\\
% ...
$^{154}$ Raymond and Beverly Sackler School of Physics and Astronomy, Tel Aviv University, Tel Aviv, Israel\\
% ...
$^{159}$ Department of Physics, University of Toronto, Toronto ON, Canada\\
% ...
\end{flushleft}


\end{document}